\documentclass[11pt]{article}

\usepackage{graphicx}
\usepackage{epsfig}
\usepackage{amsmath,amssymb,amsthm,bm}
\usepackage{color}
\usepackage{subcaption}
\usepackage[margin=1in]{geometry}

\def\figDir{Figures}

\def\bitem{\begin{itemize}}
\def\eitem{\end{itemize}}

\def\bnum{\begin{enumerate}}
\def\enum{\end{enumerate}}
\def\beqn{\begin{equation}}
\def\eeqn{\end{equation}}

\def\eps{\epsilon}

\newcommand{\myf}{\mathbf{f}}
\newcommand{\myF}{\mathbf{f}}
\newcommand{\myg}{\mathbf{g}}
\newcommand{\myG}{\mathbf{g}}
\newcommand{\mya}{\alpha}
\newcommand{\myb}{\beta}
\newcommand{\maxG}{g^{\max}}
\newcommand{\balph}{\bm{\alpha}}
\newcommand{\bbet}{\bm{\beta}}
\newcommand{\CC}{{\cal C}}
\newcommand{\fC}{{\mathfrak{C}}}

\def\RR{\mathbb{R}}

\def\cI{ {\cal I} }

\def\cR{ {\cal R} }

\def\half{\frac{1}{2}}

\def\bpMat{\begin{pmatrix}}
\def\epMat{\end{pmatrix}}

\title{Large Scale Probabilistic Simulation of Renewables Production}
\author{Mike Ludkovski\thanks{Department of Statistics and Applied Probability, University of California Santa Barbara, USA, 93106-3110; email: ludkovski@pstat.ucsb.edu ; $^\dag$Scoville Risk Partners, 101 Carnegie Center Dr Princeton, NJ, USA, 08540}
\and
 Glen Swindle$^\dag$ \and Eric Grannan$^\dag$
}

\date{\today}

\begin{document}
\maketitle

\begin{abstract}
We develop a probabilistic framework for joint simulation of short-term electricity generation from renewable assets. In this paper we describe a method for producing hourly day-ahead scenarios of generated power at grid-scale across hundreds of assets. These scenarios are conditional on specified forecasts and yield a full uncertainty quantification both at the marginal asset-level and across asset collections. Our simulation pipeline first applies asset calibration to normalize hourly, daily and seasonal generation profiles, and to Gaussianize the forecast--actuals distribution. We then develop a novel clustering approach to stably estimate the covariance matrix across assets; clustering is done hierarchically to achieve scalability. An extended case study using an ERCOT-like system with over 500 solar and wind farms is used for illustration.
\end{abstract}

\section{Introduction}

The stochastic nature of renewable energy generation (primarily understood to be grid-scale solar and wind farms) necessitates probabilistic analysis of future generation which is in turn used as an input for unit commitment and economic dispatch decisions. For example, reliability assessment and the need for reserves depend critically on potential deviations of actual generation from forecasted levels. To do so, one must be able to sample scenarios of realized generation obtained from a generative model, rather than, say, from bootstrapping a fixed set of weather scenarios arising from a numerical weather forecast.

In this article we tackle the problem of developing a stochastic model for generating system-wide scenarios of future renewables generation. Here we are concerned with the daily and hourly timescales, the prototypical context being day-ahead unit commitment that is performed by regional transmission operators. As a motivating example, given an asset-level forecast for the next 24 hours, our goal is to sample actual energy production across many renewable assets, again at the hourly scale. The typical setting for deployment of this platform involves several hundred generation assets of two or more types.


The key feature of our approach is the ability to construct a \emph{joint} model across all the assets.  Thus, each outputted simulation scenario is a large multivariate matrix, with rows indexing assets and columns indexing the 24 hours. These simulations are intended to be consistent both with single-asset behavior (marginals) as observed historically, as well as observed correlations between assets; the latter includes capturing correlation across asset types (solar vs.~wind) and zonal laws.
Accurately capturing the respective dependence across assets and time is the crux of our contribution.

The other fundamental challenge that we tackle is the relative paucity of historical data. One typically has access to a year or two of relevant forecasts/actuals, and as a result there are substantial limitations to what statistical methods are applicable to calibrate the high-dimensional correlation structure. For example, in the context of unit commitment where the delivery time scale is hourly, the number of random variables being simulated for a single delivery day in the ERCOT case study used for illustration below is $\approx$500 assets $\times$ 24 hours  $\approx 10^4$. The data available for calibration is typically $\ll 1000$ days of historical forecast and actual volumetric data. Given the seasonal variation in the behavior of the asset types under consideration, the data relevant for calibration for a particular delivery day is a fraction (say 25\%) of the total available data. This renders nontrivial both characterizing the dynamics of each individual asset and estimating correlations between the assets. Indeed, with the length of the time-series $p$ roughly matching the number of assets $J$, direct inference of the $J \times J$ covariance matrix is ill-posed. To this end, we propose a hierarchical approach that recursively breaks the problem into estimating smaller $J_\ell \times J_\ell$ matrices and then reassembling these into a block structure.

Beyond calibrating the cross-asset correlation, the context of renewable energy simulations imposes several other demands on the stochastic engine. First, simulated power levels must not only respect the non-negativity constraint, but also satisfy the hour-, day- and asset-specific maxima. These are especially pronounced for solar assets, where the daily and annual cycles of solar radiation create multiple layers of seasonality.  Second, the simulated scenarios must capture this mixed distribution, which becomes especially important when considering aggregated production across multiple assets. Indeed, the 
distribution of realized generation exhibits significant point masses: there are nontrivial probabilities of both zero and maximum generation, along with a continuous distribution within those bounds. Third, because scenarios are generated \emph{conditional} on a forecast, it is critical to capture the forecast-actuals dependence, such as the conditional heteroskedasticity. Finally, our scenario assessment is driven by the impact of renewable generation on the grid, i.e.~aggregate production across several assets as reflected in the transmission network and respective security constrained unit commitment (SCUC) and economic dispatch (SCED). Therefore, the quality of the joint scenario distribution is primarily based on how well it captures aggregate production, rather than  abstract statistical metrics.

\textbf{Literature review:} 
 At the bird's-eye level, our simulation platform can be placed in the landscape of synthetic dataset frameworks, see for example \cite{golestaneh2016generation,larraneta2019generation,ramirez2021simulation,zhang2018stochastic,buster2021physical} for other ways to build synthetic solar and wind scenarios. We also refer to the competitions \cite{hong2016probabilistic,hong2019global}. To our knowledge, ours is the first platform to combine simulations across multiple asset types. The latter raises the challenge of the varying number of active hours (i.e.~correlating solar assets that are only active in the daytime with wind assets that are active throughout).

While there are hundreds of articles on forecasting renewable generation, by and large these focus on point predictions. For example, the emerging machine learning techniques are excellent in minimizing the predictive error but are not designed to quantify uncertainty around those predictions. For the probabilistic forecasts that we aim for, one may mention  \cite{lauret2019verification,woodruff2018constructing,buster2021physical} for solar generation and \cite{staid2017generating,rachunok2020assessment,li2020clustering,pinson2012evaluating,gneiting2006calibrated,mclean2013probabilistic}
for wind generation. Note that some of the above works focus on predicting horizontal global irradiance (the dominant driver of solar energy production) or wind speed (which drives wind production), while we concentrate on direct modeling of quantities measured in megawatt-hours. While applying the (nonlinear) production curve to map from weather inputs to MWh outputs is feasible for analysis of a single, fully known asset, it is extremely challenging to carry out in bulk, especially since not all asset characteristics may be publicly known. Consequently, we believe that working solely in MWh-universe is more appropriate; this choice does restrict some of the available forecasting tools. A recent overview about integration of probabilistic forecasts in grid operations is provided in
 \cite{li2020review}. 
 
Closer to the aim in our work, there is a literature strand addressing spatio-temporal probabilistic methods that simultaneously model renewable generation at multiple geographic locations. Existing approaches include copulas \cite{munkhammar2017copula,panamtash2020copula,tang2018efficient}; kriging or Gaussian processes \cite{aryaputera2015very,van2018probabilistic,yang2018power,wytock2013sparse} and downscaling of weather forecasting ensembles \cite{buster2021physical}. Among copula approaches, one may distinguish the application of vine copulas  \cite{wang2017probabilistic}, tail copulas \cite{muller2020copula} and Bayesian copulas \cite{panamtash2020copula}.

Relative to the bulk of the forecasting literature our setup has two critical distinctions. First, our primary interest is in probabilistic forecasts that provide the full joint distribution across all assets of interest. To that end, we concentrate on assessment using statistical scoring methods, rather than on minimization of predictive error.  We utilize the framework
of strictly proper scoring rules introduced by Gneiting and Raftery
 \cite{gneiting2007strictly,gneiting2007probabilistic,gneiting2014probabilistic} and summarized for our context in Section \ref{sec:assess}. Gneiting already applied these to wind speed analysis in \cite{gneiting2006calibrated}; for an updated overview in the context of renewable generation see  \cite{lauret2019verification}. Related works that assess probabilistic forecasts of renewables are in
  \cite{woodruff2018constructing,rachunok2020assessment,ziel2018probabilistic}.

Second, our motivating application is the sampling of day-ahead hourly scenarios conditioned by a respective forecast. Such  Numerical Weather Predictions (NWP) are received daily by the market participants and system operators and drive the downstream tasks of unit commitment, economic dispatch, and risk measurement. This day-ahead setup implies that we view all quantities as vectors, indexed by (active) hours and collected daily, rather than as time-series. Moreover, the non-trivial dependence between actuals and forecasts implies that looking at forecast errors on their own is insufficient and that the dependence between conditioned realizations materially differs from the unconditional one.  For example, the tail dependence between conditioned forecasts is much weaker than between unconditional ones.



The rest of the article is organized as follows. Section \ref{sec:platform} summarizes our platform and the data that we work with. Section \ref{sec:clustering} describes our clustering algorithm which is the main methodological contribution. Section \ref{sec:calibration} describes the de-trending, rescaling and calibration applied prior to the clustering and then their application in reverse to generate scenarios given the fitted covariance structure. Section \ref{sec:case-study} presents a case study for a large system with 200+ each of solar and wind farms; Section \ref{sec:conclude} concludes.

\section{Platform Overview and Data}\label{sec:platform}

\subsection{Process Flow}

\begin{figure}[tbph]
\begin{center}
\includegraphics[width=6in,trim=0.6in 3.8in 0.6in 0.65in,clip=TRUE]{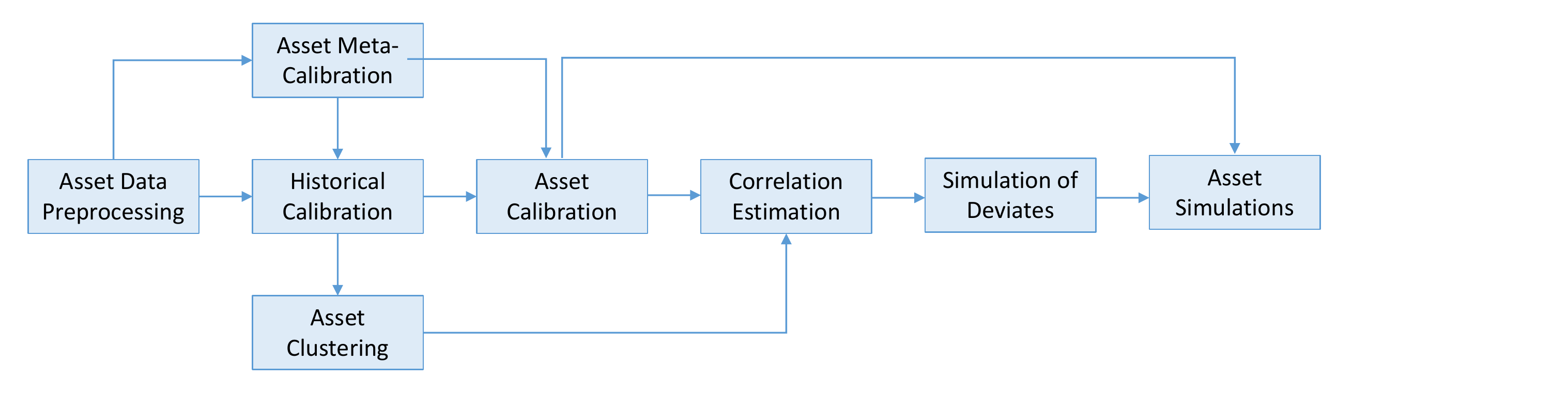}
\caption{{\label{FlowChart} Workflow of the simulation platform.}}
\end{center}
\end{figure}

At a high level, the proposed method is summarized in Figure \ref{FlowChart} and conceptually involves (A) calibration, namely standardizing and normalizing raw data to render it Gaussian-like, whereby we extract the respective $z$-scores; (B) hierarchical clustering based on correlation-driven annealing; (C) simulation based on conditional Gaussian draws, which is then fed in reverse order to the calibration module.


The platform accepts as input historical forecasted and actual power generation for each asset. The following modules are then applied:
\bitem
	\item[(i)] Asset Data Preprocessing: Organization of historical actual and forecasted volumes as well as metadata.
	\item[(ii)] Asset Meta-Calibration: Parameterization of key operational attributes, notably maximum production and diurnal envelopes. This is separated from generic daily calibration due to the global nature of the attributes being parameterized and the associated time required.
	\item[(iii)] Historical Calibration: Model calibration by asset for a set of dates spanning the historical date range---usually this is all dates. This yields a set of normal deviates for each asset associated with each hour of delivery and spanning the calibration date range.
	\item[(iv)] Asset Clustering: Construction of hierarchical clusters used in the correlation representation using the results from step (iii).
	\item[(v)] Asset Calibration: Model calibration by asset for a specific simulation date---this could be in the historical date range for backtesting or for a future date.
	\item[(vi)] Correlation Estimation: Construction of intra-cluster correlations and propagation of specific asset normal deviates up the clustering hierarchy.
	\item[(vii)] Simulation of Deviates: Propagation of simulations down the cluster hierarchy with conditional normal calculations completing normal deviate simulations within each cluster.
	\item[(viii)] Asset Simulations: Conversion of normal deviates to production volumes using asset calibration results, including production and diurnal envelopes.
\eitem

Note that after steps (i)-(v), asset dynamics are represented by a set of historical normal deviates spanning the set of ``active hours" for the asset. For wind the active hours are 1:24; for solar assets active hours vary by asset and time of year creating a nontrivial impediment for most correlation estimation procedures which is the key motivation for the proposed approach in steps (vi)-(vii).

\bigskip

\textbf{Notation:}

\bigskip 

\begin{tabular}{lr|lr}
$J$ & Number of assets & $i,j$ & asset indices \\
$d=1,\ldots,D;h$ & Days, Hours & $t$ &  generic time index \\
$ \myg_d=(g_{d,h}) $ & Actual production MWh & $\mya_{d,h}$ & actuals fraction $\in [0,1]$ \\
$ \myf_d = (f_{d,h})$ & Forecast production MWh & $\myb_{d,h}$ & forecast fraction $\in [0,1]$ \\
$ \maxG_d$ & Estimated daily max & $\maxG_{d,h}$ & Estimated hourly max \\
$ \mu_h(\beta)$ & Conditional mean of $\mya$ & $\sigma^2_h(\beta)$ & cond variance of $\mya$ \\
$k=1,\ldots,K$ & index for factors & $\psi_k, \gamma_k$ & PCA factors \\
$A$ & corr matrix & $\rho_{ij}$ & entries of $A$ \\
$\fC$ & cluster & $c(\mathfrak{C})$ & delegate of $\fC$  \\
$\mathcal{C}$ & clustering hierarchy & $l=1,\ldots,L$ & hierarchy levels \\
$\eta$ & annealing param & $\ell$ & annealing rounds  \\
$E$ & sim annealing energy & $T_c$ & annealing schedule \\
$|\fC|$  &cluster size & $\kappa$ & cluster size penalty \\
$\phi$ & year frac & $\Theta$ & window width \\
$I_d = [s_d, t_d]$ & observed diurnal boundaries & $[\hat{s}_d, \hat{t}_d]$ & estimated boundary \\
$\mathcal{I}_d$ & daily window & $H_d(\cdot)$ & daily rescaling function \\
$G_{nom}$ & nominal capacity  &  & \\

\end{tabular}

\bitem
    \item There are $J$ assets, indexed by $i,j=1,\ldots,J$;
	\item Days are indexed by $d=1,2,\ldots, 365$;
    \item Hours are indexed by $h=1,2,\ldots, 24$;
    \item Generic time index is $t=1,\ldots,T$;
    \item To define distances between days, we use the corresponding year fraction $\phi(d) \in [0,1]$;
    \item For a given day $d$, we use a time window of width $\Theta$: $\mathcal{I}_d = \{ d' : |\phi(d)-\phi(d')| \le \Theta \}$;
	\item Actual generation is $\myg_d=(g_{d,h})$ on day $d$---a vector of dimension 288 (the number of 5 minute intervals in a day).  When averaged to hourly intervals, actual quantities are denoted by $\myG_d=\myG_{d,1}, \ldots, \myG_{d,24}$. Similarly $\myf_d=(f_{d,h})$  denotes forecasted (hourly) generation. Both $\myG_d$ and $\myF_d$ are $\in \RR^{24}$.
	\item Daily diurnal production boundaries pertain to solar assets. Specifically, $I_d \equiv [s_d,t_d]$; $s_d$ denotes the first interval (counted in 5-min units) of the day with positive production; $t_d$ the last. These are historical realizations for each day. The corresponding estimated diurnal boundaries are $\hat{I}_d = [\hat{s}_d, \hat{t}_d]$.
    \item Normalized generation ratios are $\balph_d \in [0,1]$ and $\bbet_d$ which correspond to $\myG_d, \myF_d$;
    \item PCA factors are $\psi_k, k=1,\ldots,$ with respective amplitudes $\gamma^j_k(d)$ indexed by days;
    \item The correlation of $\gamma_1^i(\cdot)$ and $\gamma_1^j(\cdot)$  is denoted by $\rho_{ij}$. Below for a target date $d$, we compute correlation over a window $\mathcal{I}_d$ so technically the correlation is $\rho_{ij}(d)$. The resulting covariance matrix is $A$.
\eitem

\subsection{Case Studies}





Our test set spans the ERCOT market in Texas and includes both solar PV and wind generators. The dataset was primarily created by NREL through re-analysis of numerical weather simulations and covers the 2017 and 2018 calendar years. NREL re-analyzed ensembles of ECMWF (European Centre for Medium-Range Weather Forecasts) models in order to extract predictions of solar irradiance and wind speeds at 10km spatial resolution and 5 min frequency. Interpolation was then applied to obtain respective physical quantities at the sites of the solar and wind farms. Next, the weather data was combined with asset characteristics and the respective power transfer curves to derive energy output via the WIND \cite{draxl2015wind} and SIND \cite{sengupta2018national,feng2019opensolar} toolkits. This type of re-analysis can be done both for existing assets, as well as proposed (i.e.~``synthetic") projects, permitting study of high renewable-penetration scenarios \cite{rossol2018nrel}. We note that these weather models do in fact provide  ensemble forecasts; however the ensemble dispersion is based on a different notion of uncertainty and requires separate pre-processing to yield calibrated range of actuals; in this work we therefore do not utilize ensembles and use only the mean point forecasts.

\begin{figure}
\begin{center}
  \includegraphics[height=2.3in,trim=0in 0.5in 1in 0.5in]{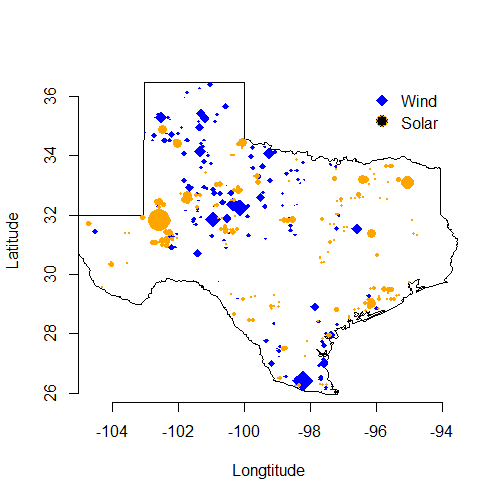}
  \includegraphics[height=2.3in,trim=0.7in 3.5in 0.7in 3.2in]{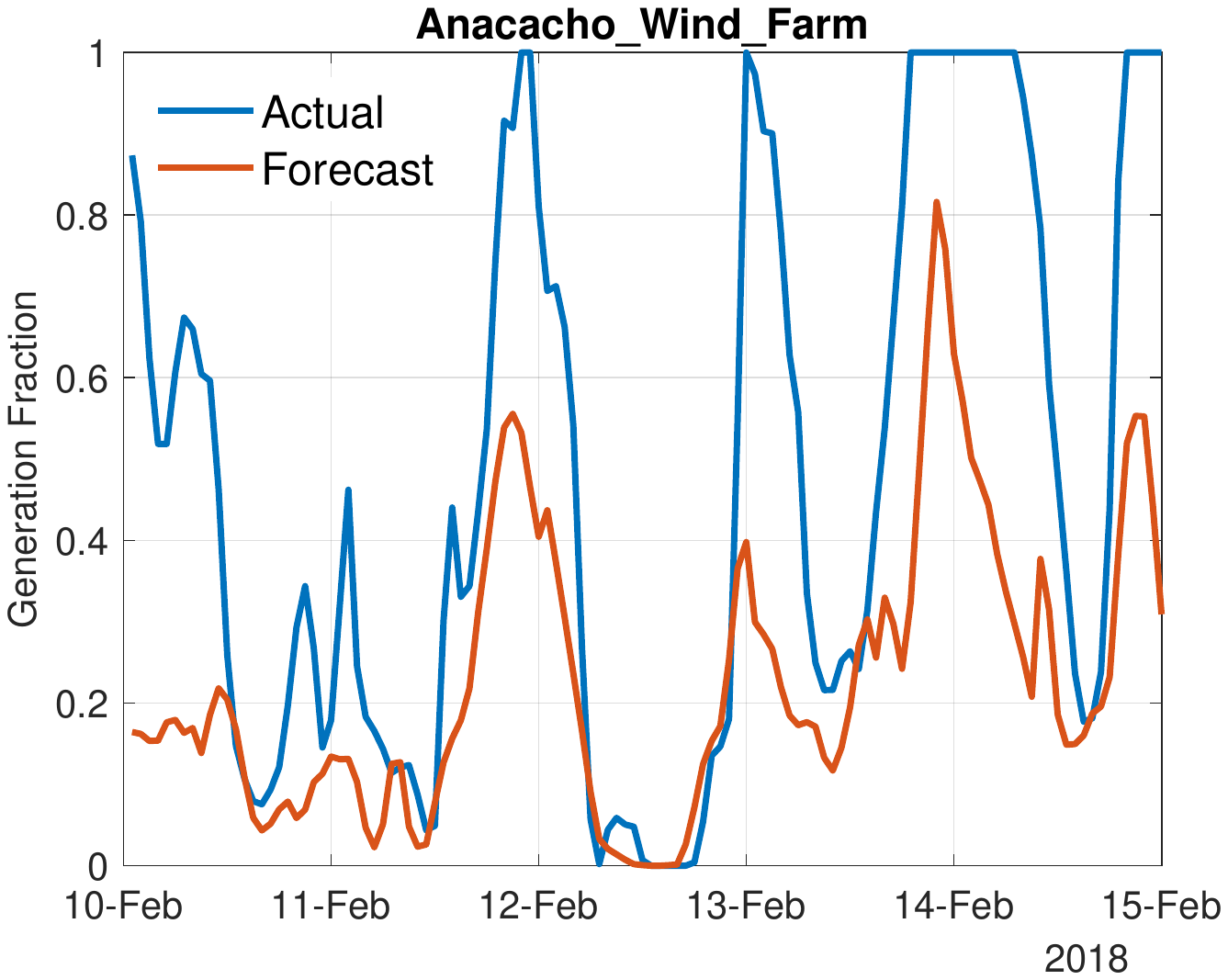}
\end{center}
  \caption{Left: Proposed renewable assets in ERCOT region. Symbol size is proportional to nominal generation capacity $G_{nom}$. Right: time series of normalized generation $\balph_{d,h}$ at Anacacho wind farm over 5 days (120 observations) in February 2018. \label{fig:ercot}}
\end{figure}

The resulting ERCOT-wide dataset that we consider consists of two testbeds: (i) \texttt{Existing}, which includes 22 solar farms and 125 wind farms; (ii) \texttt{Proposed}, which adds more than 300 additional generators, for a total of 226 solar and 264 wind assets. The projects are throughout ERCOT, although \texttt{Existing} assets occur only in some of the ERCOT zones (for example Coast for wind contains no assets). ERCOT consists of 8 zones (Coast, West, Far West, North, North Central, East, Southern and South Central), which will be used later for aggregate analysis in Section \ref{sec:case-study}. The left panel of Figure~\ref{fig:ercot} shows the locations of the 490 assets in the \texttt{Proposed} testbed. Note that nominal capacities span several orders of magnitude from 1.05 MW to 1219 MW, with a median of approximately 150 MW and total nameplate capacity of 41GW  for solar and 60GW for wind.

In total, the dataset spans $T=730$ days or 17520 hours. Given the limited size of the dataset, we utilize all of it for training and concentrate on in-sample testing. 


\subsection{Stylized Features of the Data}

Due to the widely varying nameplate capacity $G^j_{nom}$ and the dominant diurnal pattern in solar generation, analysis of raw produced MWh is not recommended due to obvious non-stationarity. Instead we normalize by de-trending and rescaling (see Sec.~\ref{sec:rescaling}) to obtain production ratios that are always in the unit  interval $[0,1]$. This normalization is further combined with a shift-and-stretch transformation to yield data that are statistically i.i.d.~across assets and days for the purposes of correlation estimation.

\begin{figure}[tbph]
\begin{center}
\includegraphics[height=2in,trim=0.8in 3in 0.8in 3.15in, clip=TRUE]{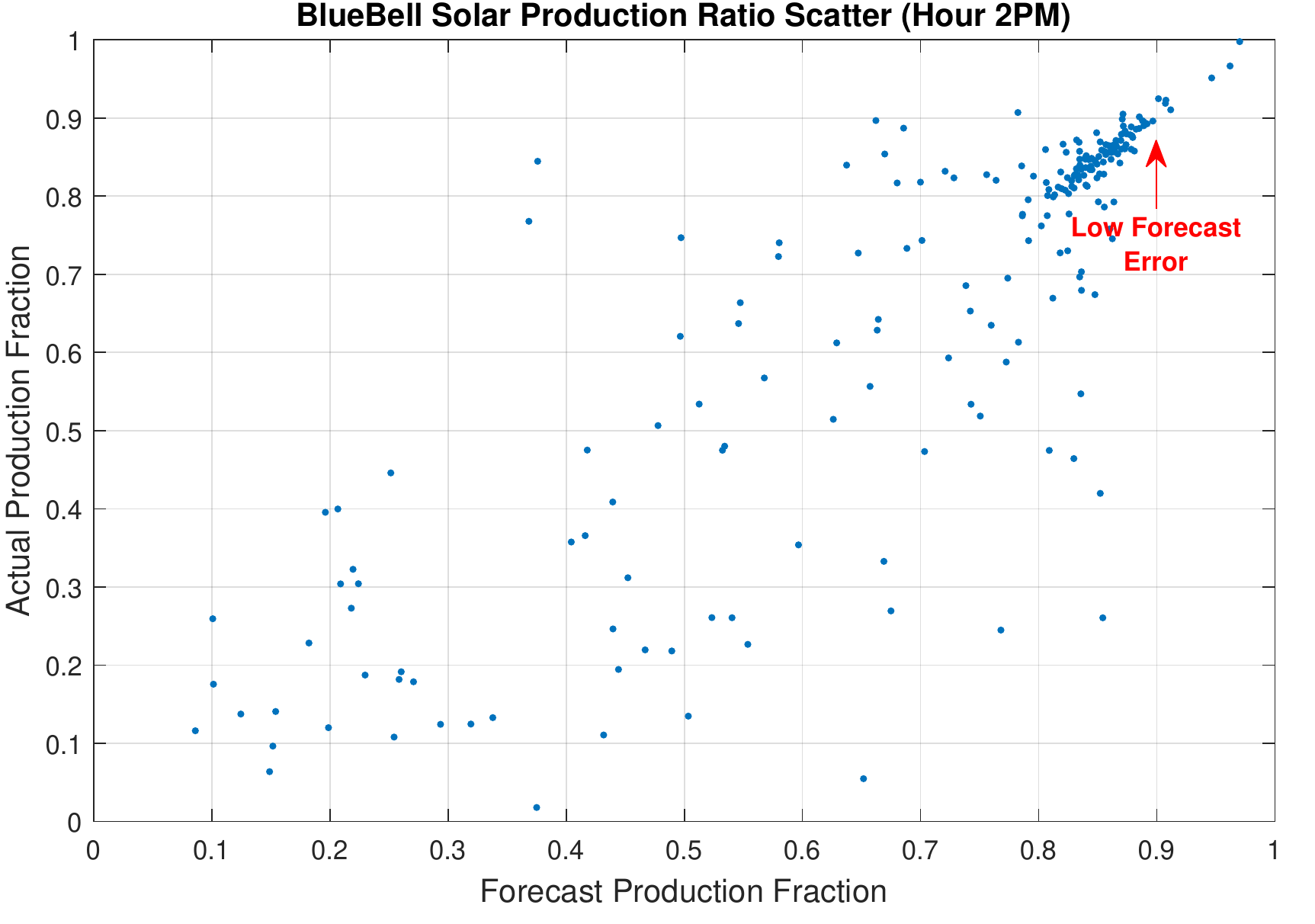}
\includegraphics[height=2in,trim=0.8in 3in 0.8in 3.15in, clip=TRUE]{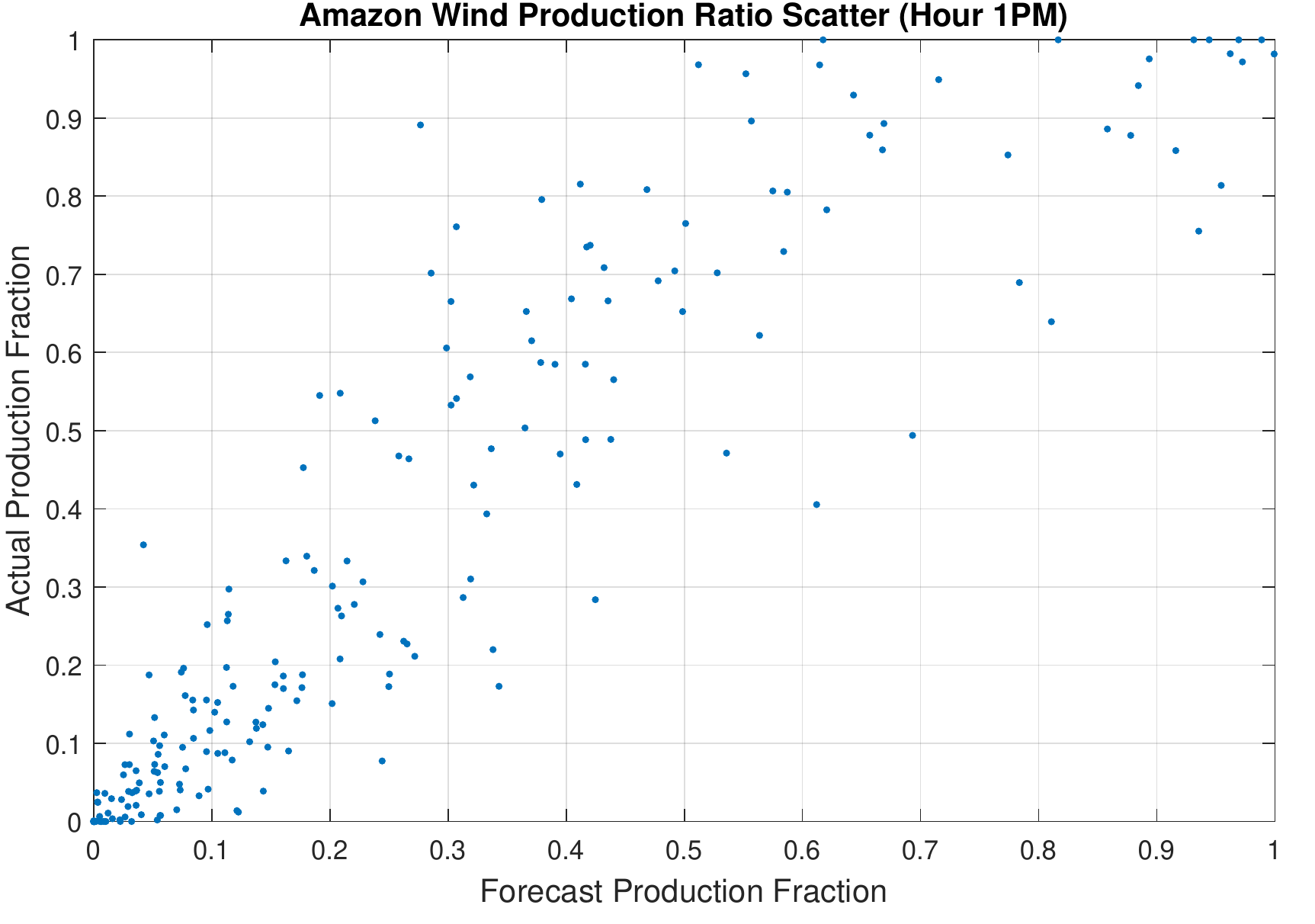}  
\caption{\label{fig:stylized} Left: actual $\alpha_{\cdot,h}$ vs forecast $\beta_{\cdot,h}$ production ratios for a solar asset (Blue Bell) at $h= $2pm.  Right: same for a wind asset (Amazon Wind Farm) at $h=$1pm and a window of 218 days.}
\end{center}
\end{figure}

A production ratio of zero means that no power is produced and 1 means that maximum possible production (given hour, day, and asset capacity) is achieved.  The right panel of Figure~\ref{fig:ercot} displays a sample hourly time series for a wind asset forecast and actuals, $\alpha_{d,h},\beta_{d,h}$. The key feature is the nontrivial number of zeros and ones in the actuals $\alpha_{d,h}$ ---on many hours either there is zero wind and the turbine is not spinning, or the turbine spins at its maximum rate and realized generation fraction is 1. For wind assets, there are about 4-7\% of hours with zero generation, and about the same for max-generation. Additional daily patterns are manifest. There is much less wind during the day, so for example at noon, on 8-15\% of days there is no wind energy produced at all (see mid-day of Feb 12). Similarly, strong wind primarily happens at night and at midnight 15-30\% of observed generation is at maximum capacity (see the nights of Feb 14 and 15 in  Figure \ref{fig:ercot}). As could be expected, forecasts tend to smooth out such point masses and hence are generally strictly inside  the $(0,1)$ interval. In line with these features,  joint simulation of realized volumes conditioned on forecasts must consider and account for the following:
\bitem
	\item The distribution of realized forecast errors $\epsilon_{d,h}$ (the difference between actual and forecasted ratios) being dependent upon the forecast value. Such conditional dependence is particularly striking for solar assets. The left panel of Figure \ref{fig:stylized} shows the realized fraction of maximum production, $\alpha_{d,h}$,  for a solar asset scattered against the forecasted ratio $\beta_{d,h}$. The variance of forecast error $\epsilon_{d,h}$ is noticeably smaller at high forecasted levels ---a manifestation of the uncertainty associated with potentially cloudy days, in contrast to relative certainty of a sunny forecast. Similarly, the right panel of Figure \ref{fig:stylized} scatters actuals against forecasted ratios for a wind asset ---we observe low uncertainty on windless days ($\beta_{d,h} \simeq 0$) and increasing uncertainty as wind forecasts pick up.

	\item Solar and wind asset dynamics are fundamentally non-stationary.   For example, for solar assets 8AM in January is very different than 8AM in March. Rescaling needs to account for the maximal achievable capacity varying by hour and the fact that solar start/end periods are driven by the annual sunrise-sunset cycle. Similarly, there are seasonal and diurnal patterns in average wind generation and the respective variance. Consequently, fusing observations from different days in order to estimate trends and correlations requires multiple layers of de-trending and rescaling.
\eitem


As we are interested in directly modeling electricity production, there are hard physical constraints on the simulated quantities. Assets have nominal capacity $G_{nom}$, but maximum achievable capacity $\maxG_{d,h}$ varies by hour, especially for solar, where $\maxG_{d,h}$ depends on the angle of the sun at that instant. In our setup, $\maxG_{d,h}$ is not available directly and is rather statistically inferred from the data, see Section~\ref{sec:rescaling}.

\section{Cluster Analysis for Correlation Structure}\label{sec:clustering}

In this section we describe our procedure for inferring the conditional correlations of production ratios $\balph^i$ and $\balph^j$ given the respective forecasts.
The high-dimensional nature of the problem coupled with the limited amount of historical data available for calibration means that some assumptions on the structure of the problem, particularly of the correlations between asset-delivery hours must be made. These are detailed below, but roughly speaking the approach is to construct a hierarchical relationship between asset clusters, thereby preserving high correlations between assets that have historically exhibited such behavior.

Our approach is based on constructing clusters that capture closely-related assets, and then estimating the historical correlation matrix for each cluster. This is done recursively through selecting cluster ``delegates" that are used for correlating at the parent level. Ultimately, cluster membership is used to build a structured correlation matrix that is fed into a multivariate-Gaussian framework. To this end, we address the questions of (i) defining the concept of asset similarity that underlies clustering; (ii) cluster construction; (iii) estimation of intra- and inter-cluster correlations.

\begin{figure}[tbph]
\begin{center}
\includegraphics[height=2.7 in,trim=0.8in 3in 0.8in 3in,clip=TRUE]{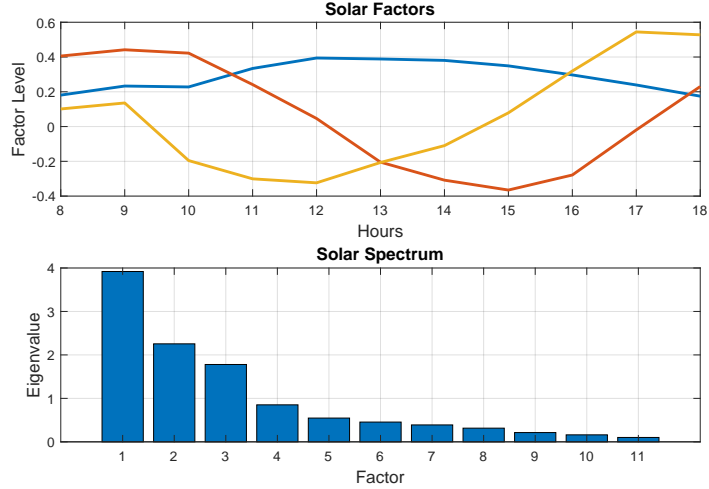} 
\caption{{\label{SolarPCA} {Sample factor analysis for a solar asset (Blue Bell). Factors $\psi_k(h)$ (top) for $k=1,2,3$ (in blue, red, orange, respectively), and eigenvalues $\lambda_k$ (bottom). The calibration is to 1/1/2018.}}}
\end{center}
\end{figure}

{\it Factor Representation:} The approach used in all correlation and clustering that follows is based upon a principal component analysis (PCA) representation of each asset's standardized/normalized deviates. The latter are obtained from a joint model for $(\alpha_{\cdot,h},\beta_{\cdot,h})$ visualized above and explained in Section \ref{sec:calibration}.
Figure \ref{SolarPCA} shows the first few eigenvalues and spectrum for a particular solar asset on January 1, namely the shape of the first 3 PCA factors $h \mapsto \psi_k(h)$ (note that $h \in \{8,18\}$ as the other hours feature zero solar production with probability 1, meaning there are only 11 factors in total) and the corresponding eigenvalues $\lambda_k, k=1,\ldots,11$. The typical empirical factor is manifest; the decay in the spectrum is useful.

The interpretation of PCA factors matches the application of factor analysis in other situations. The first factor $\psi_1(\cdot)$ (blue) is qualitatively similar across all assets and corresponds to production values higher than expected conditional on the forecast. This statement does not imply a linear response. A higher loading $\gamma_1$ of the first factor propagates back through the copula transformations that rendered the residuals normal and on which the PCA was performed. However, monotonicity is preserved and increasing the loading on the first factor increases realized production. In a similar way, increasing the second factor loading $\gamma_2$ increases production in the first active hours and decreases production in the later hours (red curve in Figure~\ref{SolarPCA}). 

All intra-asset correlation analysis is done in the above ``factor space'' rendering the joint analysis of assets with different sets of active hours easily implementable. This accomplishes two things. First, it translates the functional correlation of the curves (rendered as vectors) $\balph^i, \balph^j$ into the more interpretable correlation of the individual amplitudes $\gamma_k^i(\cdot),\gamma^j_k(\cdot)$. The latter are time-series indexed by $d$. Second, it allows us to consider correlation of vectors of different dimension, needed both to jointly model solar production (a vector of 10-15 hours) and wind production (24 hours), but also solar production in different time-zones or longtitudes/latitudes (e.g. a vector of 11 hours and a vector of 12 hours). The common features of factors $\psi_k$ across assets cease to be observed beyond the first few factors, and motivates our clustering on $\gamma_1$ as $\psi_1$ has an unambiguous sign across assets.

\subsection{Clustering via Simulated Annealing}

The key premise of the developed correlation structure is the assumption of a hierarchical form; one based upon a correlation-clustering of lower level asset clusters. Clustering is typically implemented  by asset type, i.e.~separate between solar and wind. The top level intra-asset clusters are handled specially as described below.
\bitem
	\item[-] All clustering currently uses the first factor amplitude, $\gamma_1$
	\item[-] At any level of the hierarchy each cluster $\fC$ is represented by the first factor amplitude of a specific ``delegate" or ``centroid" asset.
	\bitem
		\item[-] At the lowest single-asset level this is just the first factor amplitude $\gamma_1$.
		\item[-] At each of the subsequent levels the ``centroid" is the asset with the highest average correlation with other member assets.
	\eitem
	\item Clustering is via simulated annealing (SA) with an energy function of the form:
	\begin{align}\label{eq:clust-pen}
E({\cal C}) := \sum_{ \fC \in{\cal C}} \left [ 1 + \frac{\kappa}{|\fC|-1}\sum_{i,j\in \fC} \left(1-\rho_{i,j}^2\right) \right ]\end{align}
	where $\cal C$ denotes the set of clusters at step of the annealing process; $|\fC|$ is the number of elements in cluster $\fC$. $\rho_{i,j} = \mathrm{Corr}(\gamma^i_1(\cdot), \gamma^j_1(\cdot))$ is the correlation between first factor amplitudes of members $i$ and $j$. Finally, $\kappa$ is tuned dynamically to target a user-specified cluster count reduction between levels. Note that there is an implicit competition between cluster size and average intra-cluster correlation reflected in the $\kappa$-term. One way to view this penalty is the product of the size of the cluster $|\fC|$ and the average over all pairs of $1-\rho^2_{i,j}$.
\eitem

A given choice of $\kappa$ enforces a soft constraint on the achieved cluster sizes $\{ |\fC|\}$. We generally seek clusters of 2-5 assets. This implies that larger testbeds with more assets will end up with more layers of the cluster hierarchy, compared to smaller ones. For example, we obtain 3-4 layers in the smaller \texttt{Existing} testbed and 5 layers in the larger \texttt{Proposed} testbed. Smaller $\kappa$ results in flatter hierarchies.

\textbf{SA Perturbations:}
The idea of SA-based clustering is to gradually search for the best cluster assignments, namely those that minimize the energy function \eqref{eq:clust-pen}, via performing local perturbations on the cluster partition that are probabilistically explored. This is done via a loop over $\ell$ that moves from a current partition to a new one as follows:
\bitem
\item[-] Generate a candidate partition $\CC'_\ell$ by perturbing the current one $\CC_\ell$;
\item[-] Compute the corresponding change of energy $\Delta E_\ell = E(\CC'_\ell) - E(\CC_\ell)$;
\item[-] If $\Delta E_\ell <0$, accept the perturbation; otherwise, accept the perturbation with probability $\exp( -T_\ell \Delta E_\ell)$ where $T_\ell$ is the current annealing \emph{temperature parameter}. If the perturbation is rejected, then the new partition is the same as the current partition;
\item[-] Lower the temperature by a factor $\eta$: $T_{\ell+1} \rightarrow \eta T_\ell$.
\eitem

We propose two types of local perturbations:
\begin{enumerate}
  \item Merge: pick two assets $i, j$ and move $i$ to the cluster $\fC_\ell(j)$ of $j$. If $i$ was in a singleton cluster $|\fC_\ell(i)|=1$, that would reduce total number of clusters by 1;
  \item Split: pick an asset $i$ and split the cluster it belongs to, $\fC_\ell(i)$ (assuming it has more than 1 member) into 2 clusters, which would increase the total number of clusters by 1.
\end{enumerate}
Note that depending on current partition, some perturbations are ruled out (for example, splitting a singleton, or trying to add to a cluster that is already of maximal size).  The SA algorithm is initialized with every asset being in a singleton cluster and proceeds for a fixed number of $L$ steps.

Among the tuning parameters is the annealing schedule $\eta$, the number of annealing steps $L$, and the cluster penalty $\kappa$. Furthermore, we have the form of the energy function where we can perturb the original correlations $\rho_{i,j}$ in order to penalize deviations of high correlations more than those that are lower.

%
%
%
%
%
%
%
%
%
%

\begin{figure}[tbph]
\begin{center}
\includegraphics[height=2.3in,trim=0.8in 2.75in 0.7in 2.5in,clip=TRUE]{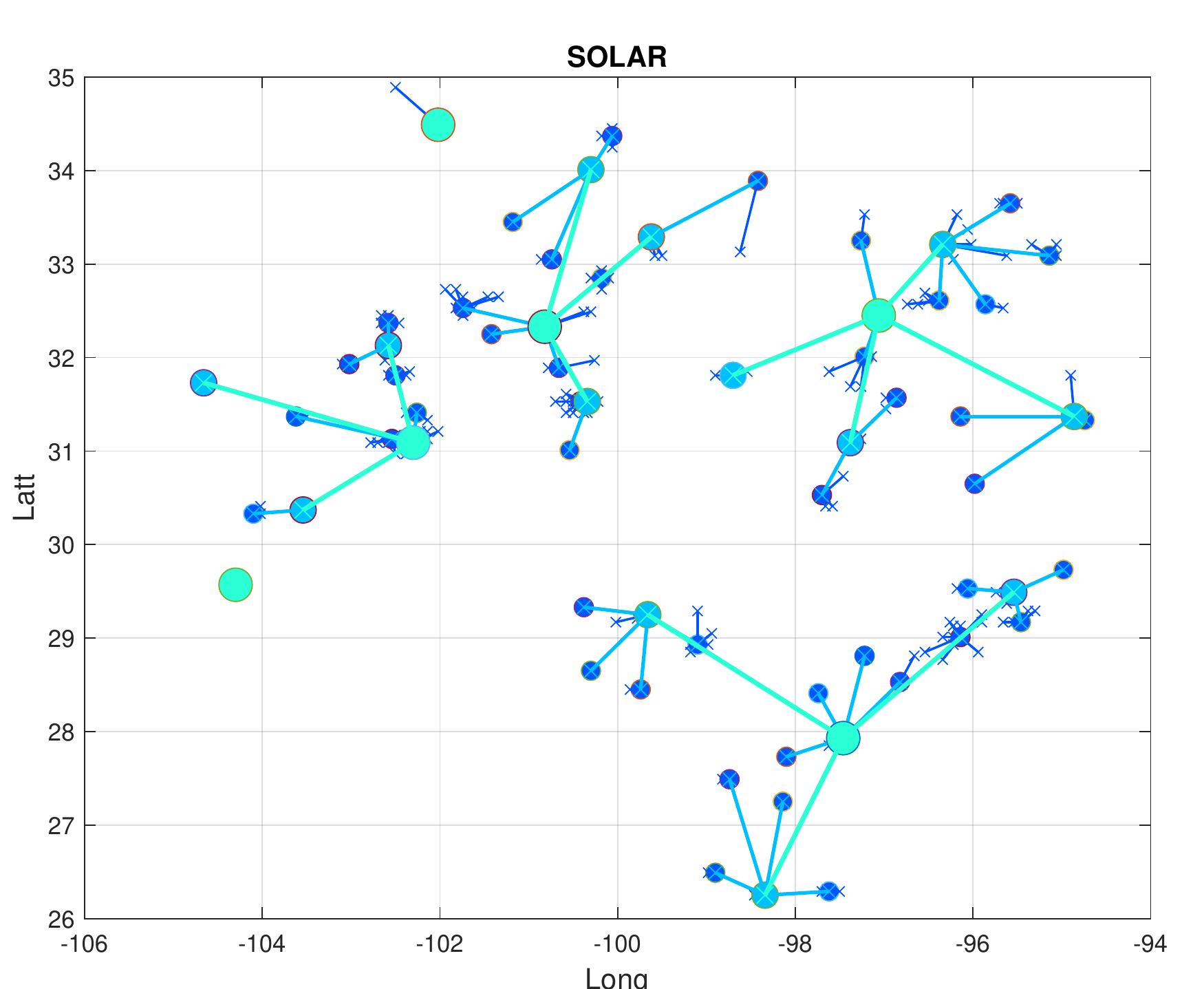}
\includegraphics[height=2.3in,trim=0.8in 2.75in 0.7in 2.5in,clip=TRUE]{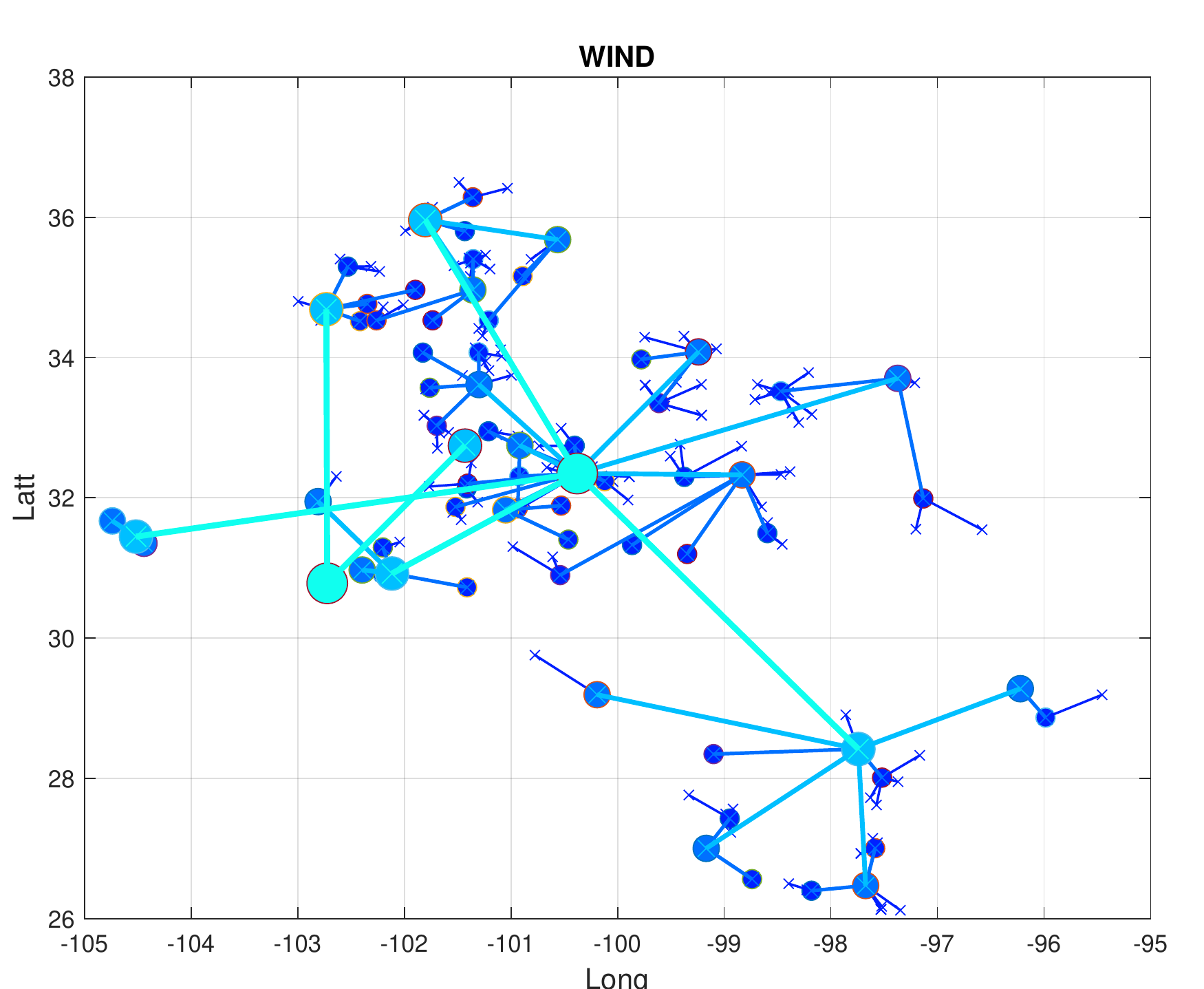}
\caption{\label{SolarLevel3} {Left: Clusters for 226 solar assets on Dec 1, 2017. Right: clusters for 264 wind assets on Apr 1, 2018.}}

\end{center}
\end{figure}

Figure \ref{SolarLevel3} visualizes the cluster hierarchy for two typical days, for both solar and wind assets. We aim to have a reduction factor of $0.3$ (see $\kappa$ above) for each level of the hierarchy.  Centroids of higher level clusters are indicated with progressively larger symbols and lines indicate cluster membership.
We observe a strong geographic contiguity of clusters across all levels of the hierarchy. In particular, the North/South and East/West divisions are clear, and for instance the wind assets in South and coastal Texas are usually in separate top-level clusters compared to the rest of the system. In this case study there are many physical agglomerations of several assets very close to each other. These are almost always clustered together at the lowest level of the hierarchy, resulting in very tight (geographically) clusters that look like a single ``blob" in Figure \ref{SolarLevel3}.

The main role of the clusters is to construct a low-dimensional correlation matrix which can stably estimate the true correlations. To this end, we do not seek a sophisticated clustering approach; moreover we do not interpret the cluster assignments as a ``hard" partitioning of the assets, but as a scaffolding that regularizes the inference of $\widehat{A}$. Thus, the plots in Figure~\ref{SolarLevel3} are primarily for illustrative purposes and do not directly reflect the quality of the clusters. In particular, there is limited practical meaning to a particular asset being designated a centroid.

\begin{figure}[tbph]
\begin{center}
\includegraphics[height=2.2in,trim=0.8in 2.75in 0.7in 2.5in,clip=TRUE]{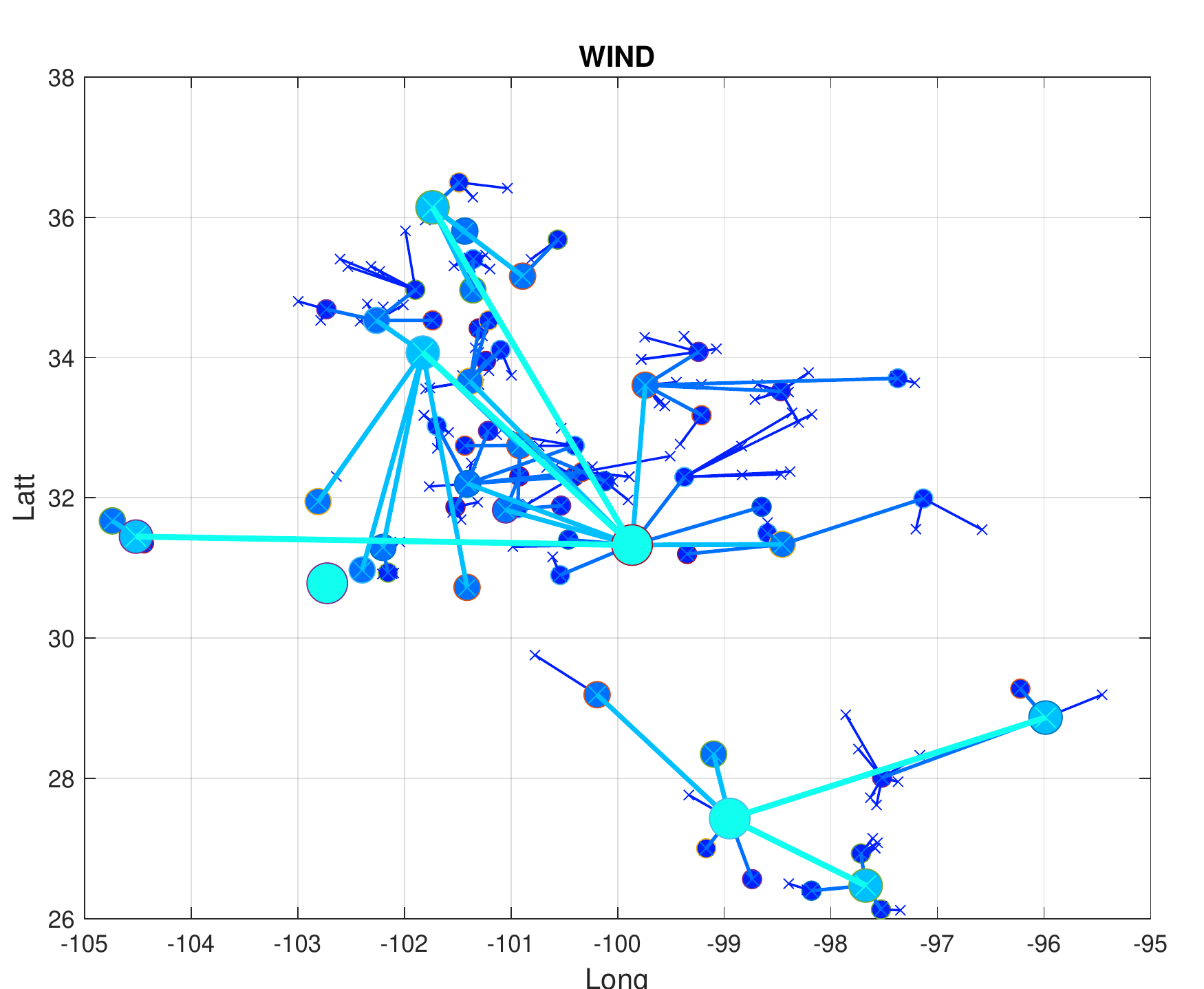}
\includegraphics[height=2.2in]{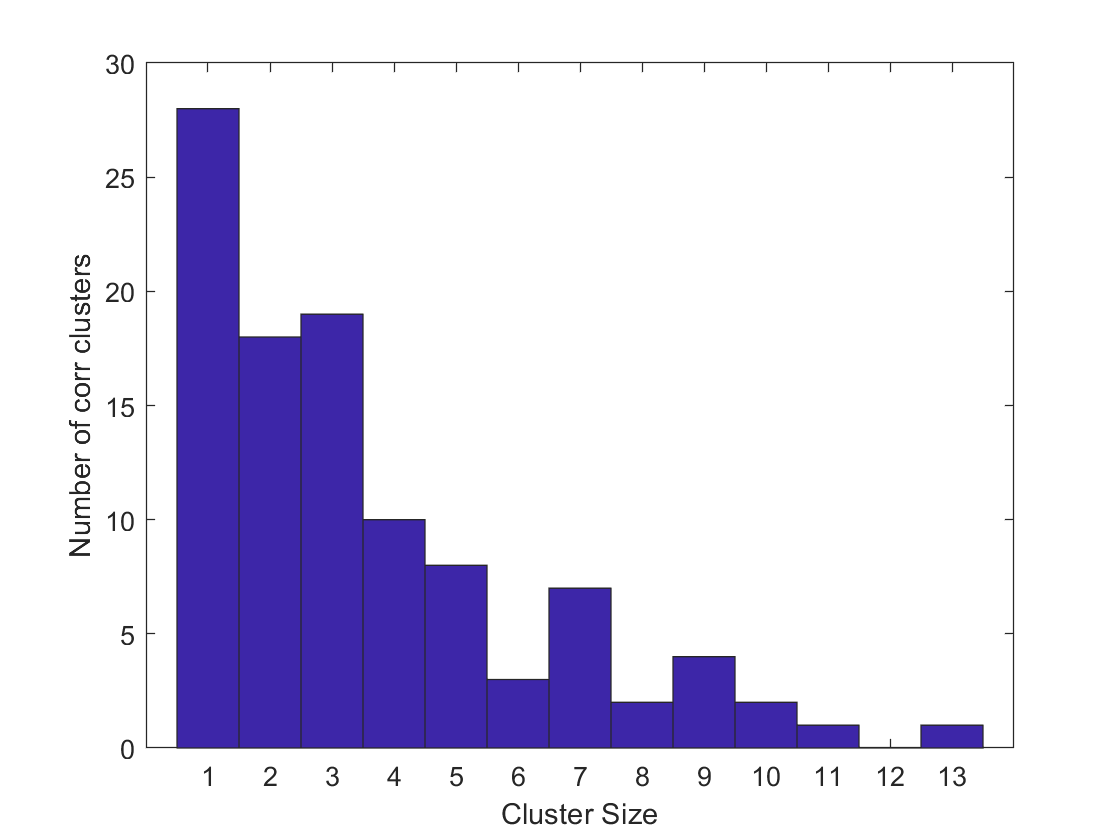}
\caption{\label{clusterHist} {Left: clusters for 264 wind assets on Feb 1, 2018. Right: distribution of $|\fC|$ on that day}}

\end{center}
\end{figure}

Figure \ref{clusterHist} shows the distribution of cluster sizes $|\fC|$, i.e.~the number of assets that constitute a cluster. By construction, the median cluster size is 3 ($\simeq 1/\kappa$), however clusters can vary substantially. About a quarter of clusters are singletons, i.e.~the annealing algorithm did not find any suitable other assets to group them with. In that particular example for Feb 1, 2018, we start with 264 distinct assets; there are then 67 clusters at level 2, 25 at level 3, 8 at level 4 and finally 3 at the top level 5, yielding a clustering tree structure with 103 clusters (nodes) at 5 levels, as shown in right panel of Figure \ref{clusterHist}.

Based on the selected parameters, with a few hundred assets we end up with 4-5 layers of cluster hierarchy. Note that by construction simulated annealing involves randomization, i.e.~running SA twice with exactly same inputs will yield slightly different results. Similarly, the user has a choice of building a clustering structure separate for each target day, or freezing the cluster assignment across a range, or possibly across the entire year. In the latter case, we recommend computing PCA factors across the date range, in order to capture the typical average behavior.


\subsection{Upward Correlation Propagation}

Any attempt to render high-dimensional correlation estimation useful and computationally tractable involves some decision about what correlations to keep. Our approach posits that:
\bitem
	\item[--] Factors within each asset class are similar in form---especially among the higher factors. More precisely, the magnitude of the inner products of factor $k$ for assets $i$ and $j$ are high.
	\item[--] Cross-factor correlations are low---the (by construction) zero correlation between different factors of the same asset extends to intra-asset: that is, the correlation between (asset,factor) $(k_1,i_1)$ and $(k_2,i_2)$ are low when $i_1\neq i_2$.
\eitem
We choose to: a) keep same-factor correlations and b) zero cross-factor correlations across assets; both  assumptions are observed empirically.

 The iteration proceeds as follows:
\bitem
	\item[-] Every cluster has a set of member assets for which the correlation matrix of factor amplitudes $\gamma_k^j$ is computed. This yields a set of correlation matrices $A^{(k)}_{L,\fC}(i,j)$ for cluster $\fC$ at hierarchy level $L$ and factor $k$ with each component $(i,j)$ corresponding to the correlation between the $k^{th}$ factor amplitude $\gamma_k$ of members $i$ and $j$.
	\item[-] A subset of a user-specified number $p$ of members of cluster $\fC$ is selected to be used at the next clustering level (recall clusters were already defined before). The set of $p$ members for upward propagation is selected to minimize the trace of the first-factor covariance matrix conditioned on the first factor amplitudes of such subsets. Essentially, the assets that persist to level $L+1$ are those that span the first-factors of the entire set most effectively.
	\item[-] Each asset type ultimately has a top level of clusters---recall that the clustering hierarchy is terminated once a prescribed cardinality is achieved (usually this is in the range of 1-4 clusters). Cross-asset correlation structure is accommodated a final covariance matrix constructed from the factor amplitudes from the first factors $\psi_k$ of each member of the top level clusters. So, for example, under the setting where each top level asset can contribute its top two factor loadings $\gamma_1, \gamma_2$ to the top-covariance, a cluster with 3 members would contribute 6 factor amplitudes. The result is a covariance matrix of dimension roughly half of the sample size in the results that we present below. We note, however, that the effective dimension  of this matrix is in the single digits, rendering covariance estimation tractable \cite{koltchinskii2017new}. We have also explored use of GLASSO methods to yield sparse precision matrices.
\eitem

\section{Calibration and Simulation}\label{sec:calibration}

{\bf  Meta-Calibration}: The purpose of the meta-calibration step is to estimate generation profiles (maximum generation for solar; mean for wind) and diurnal boundaries for solar. The purpose of these estimates is to facilitate transformation of historical production quantities at nearby dates to values consistent with the statistical attributes of a particular target date being simulated. In other words, we take a transductive approach, where the model is fundamentally based on a particular prediction set and is then locally estimated as target dates change. This choice is driven by (i) the strong seasonality at multiple levels; (ii) the non-constant dimension for the simulations, namely the varying number of active periods for solar generation both across days and potentially across different assets due to latitude effects. We refer to \cite{muller2020copula} for a related discussion of calibration in the context of modeling GHI.

\subsection{Seasonality}\label{sec:rescaling}

\begin{figure}[tbph]
\begin{center}
\includegraphics[width=0.39\textwidth,trim=0.8in 2.95in 0.8in 2.95in, clip=TRUE]{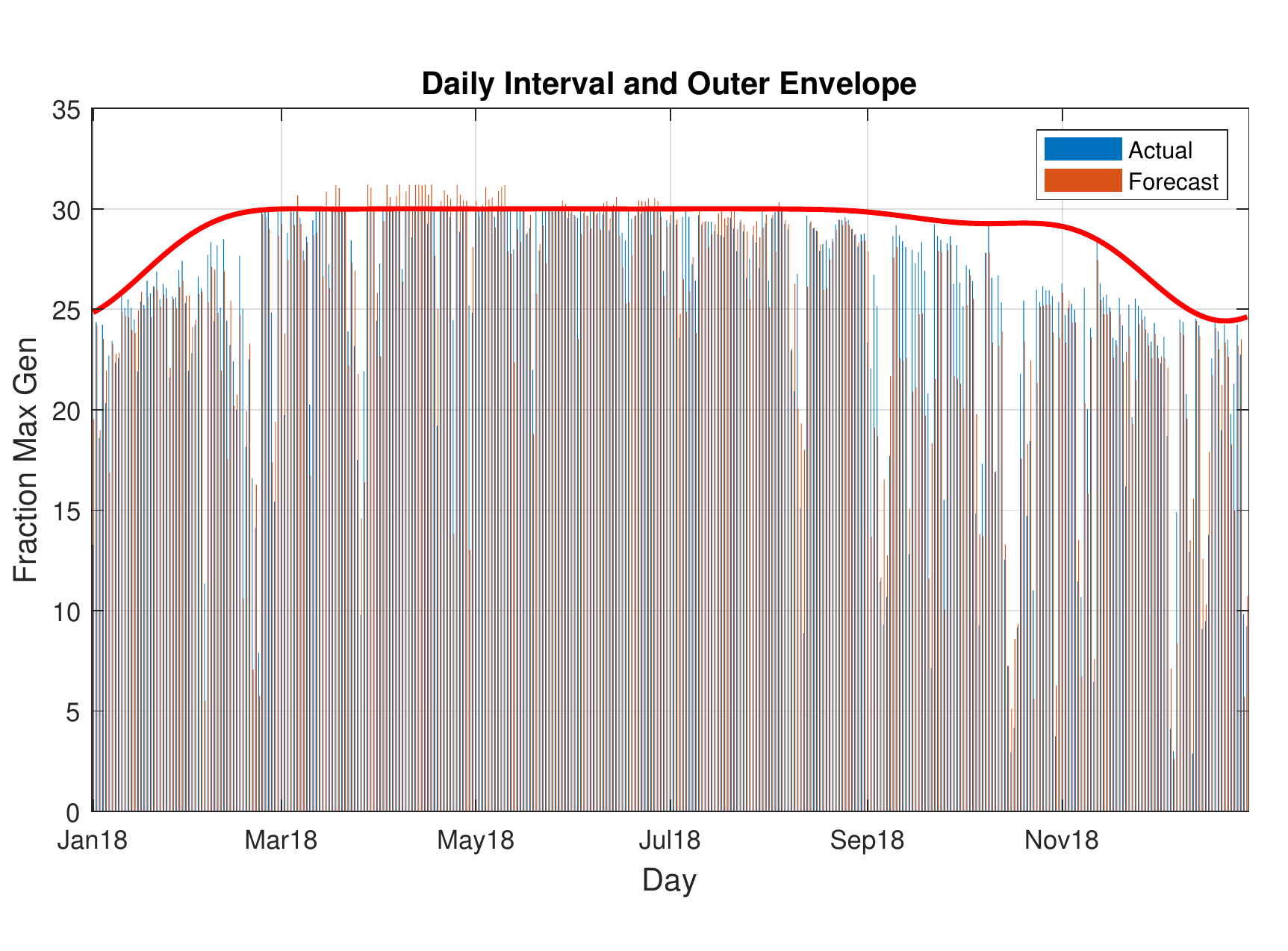}
\includegraphics[width=0.39\textwidth,trim=0.8in 2.95in 0.8in 2.95in, clip=TRUE]{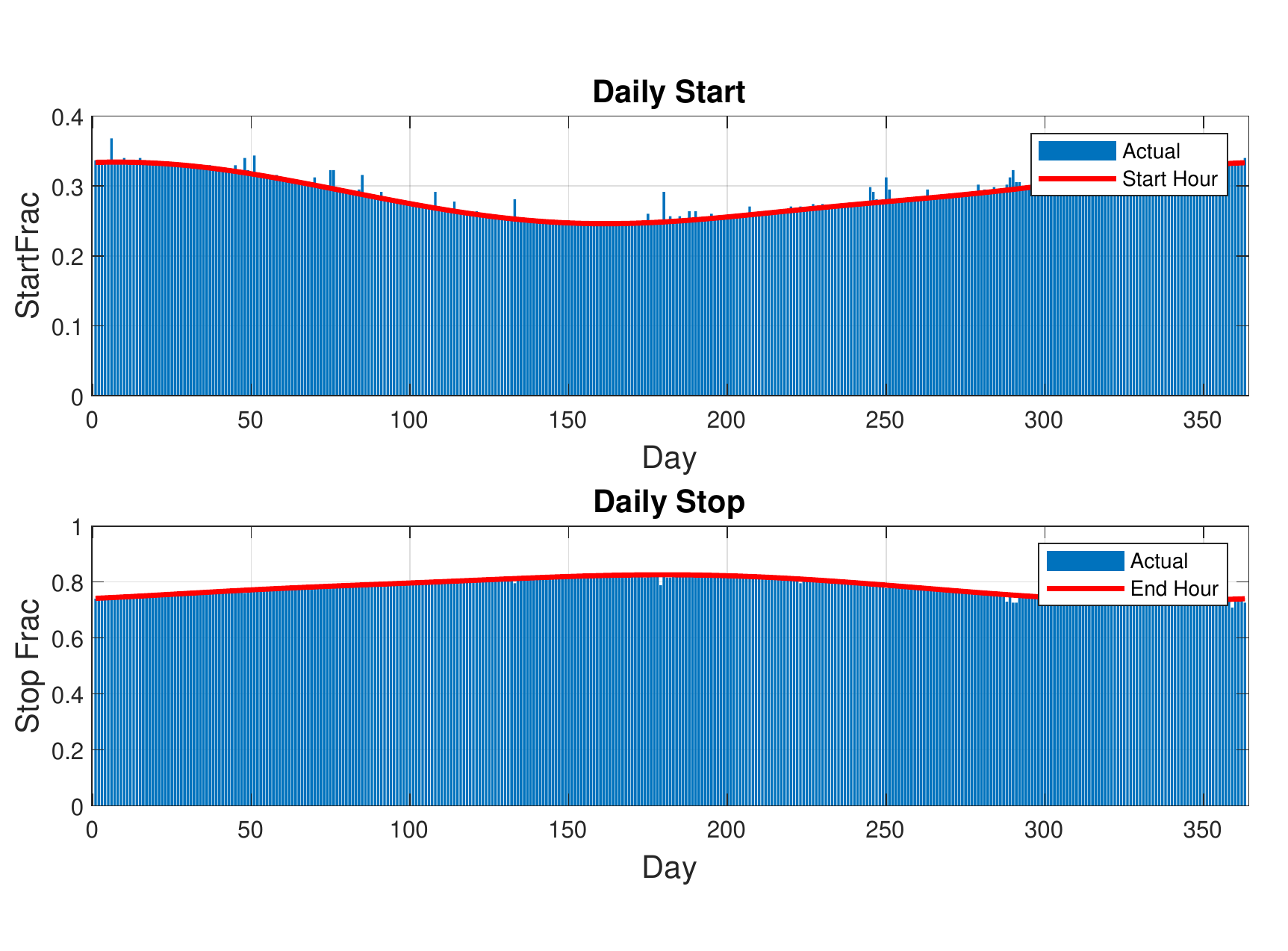}
\caption{{\label{fig:solarBnd} {Solar asset calibration. Left: maximum daily production envelope $d\mapsto \maxG_d$ for Blue Bell solar farm. Right: respective solar diurnal envelopes, $\hat{s}_d$ (top) $\hat{t}_d$ (bottom), $d=1,\ldots, 365$; (also observed $s_d, t_d$), expressed as fractions of the entire 24-hour day.}}}
\end{center}
\end{figure}

Solar and wind generation each exhibit nontrivial seasonality in production. This is visible for solar assets at the daily time-scale for maximum achievable production in the left panel of Figure~\ref{fig:solarBnd}. Similarly, shorter time-scale variations are visible in the diurnal profiles in the right panel of Figure \ref{fig:solarBnd}. The general approach to rendering a nearby date $\tilde d$ statistically consistent with a target date $d$ is to systematically rescale the observed forecast and actual production in day $\tilde d$.

 We first estimate daily maximum production levels $\maxG_d$  by minimizing the
 asymmetric error:
 \begin{align}
 \sum_d \left [ \left ( \maxG_{d} - \check{G}_d \right ) {\bf 1} _{\lbrace  \maxG_{d} - \check{G}_d > 0 \rbrace } - \kappa_{M,1} \left ( \maxG_{d} - \check{G}_d \right ) {\bf 1} _{\lbrace \maxG_d - \check{G}_d < 0 \rbrace } \right ] + \kappa_{M,2} \max_d \left ( \maxG_{d} - \check{G}_d \right ),
 \end{align}
	where $\maxG_d = \sum_{k=1}^K \left [ \alpha_{k,1} \cos \left (2\pi k \phi(d)\right) + \alpha_{k,2} \sin\left (2\pi k \phi(d)\right) \right ]$ and we take $K$ Fourier modes.
A sample   envelope of maximum achievable solar daily production, $\maxG_d$ by date is shown in the left panel of Figure~\ref{fig:solarBnd}.
  We utilize  $K=6$ Fourier modes in order to fit relatively flat intervals in the summer period. Once obtained, these envelopes are enforced on actual and forecasted quantities---namely it is assumed that there is zero probability of any generation in any hour on day $d$ in excess of $\maxG_d$.

Next we handle the  diurnal production boundaries, i.e.~the periods outside of which solar production is zero with probability 1 due to darkness. While production boundaries intuitively correspond to sunrise/sunset at the asset location, the engineering characteristics of the asset, its local topography (e.g.~mountains) and orientation of the panels prevent their direct association with any external geophysical datasets. As such, we infer the diurnal envelopes, denoted by $\left [\hat{s}_d, \hat{t}_d \right ]$,  via another minimization of an asymmetric error over the training set. For the ``stop-gen" boundary $\hat{t}_d$ this takes the form:
\begin{align}
\min_{\tilde t_d} \sum_d \left [ \left ( \tilde t_d - t_d \right ) {\bf 1} _{\lbrace \tilde t_d - t_d  > 0 \rbrace } - \kappa_{D} \left ( \tilde t_d - t_d \right ) {\bf 1} _{\lbrace \tilde t_d - t_d  < 0 \rbrace } \right ]
\end{align}
where $\tilde t_d$ is a Fourier series (with $K=3$ in our case) and $\kappa_D$ the penalty for violating the boundary. Similar estimation pertains to the ``start-gen" boundary $\hat{s}_d$. Generation is set to zero with probability 1 outside of the estimated diurnal boundaries $[\hat{s}_d, \hat{t}_d]$. The right panel of Figure \ref{fig:solarBnd} shows $\hat{s}_d$ and $\hat{t}_d$ across the year.


For wind assets, there are no diurnal boundaries and  maximum capacity is fixed throughout the year. Nevertheless, there is still a seasonal variation of production that we infer in order to rescale and standardize different days. Indeed, if May tends to be more windy that April and we wish to consider April and May data jointly, we should be scaling May production down to maintain the same relative ranks. To this end, we estimate
an hourly mean generation surface $AvW_{d,h}$  using a Fourier series representation. This surface is employed to rescale production in a fashion similar to that used for solar. Recall that our goal is to generate i.i.d.~data across all calibration days. Thus, periods of the year where production tends to be lower should be inflated to match expected production on the target day, and vice versa. The same adjustment is done across hours of the day to make them identically distributed too.

\subsection{Rescaling}\label{sec:rescaling}

Any reasonable attempt to use temporally local data to estimate asset behavior for a given target date must first transform the nearby data to be plausibly representative of the target date. The smaller the data set, the larger the temporal windows for estimation, which renders the rescaling issue more significant. Thus, the goal of the meta- and calibration stages is to obtain a normalized training set that is i.i.d. To do so, separate calibration is done for each target date and the training data is taken from a window around that date to mitigate the annual seasonality.

Rescaling production is accomplished by dilation of the diurnal boundaries and volumetric scaling by the ratio maximum production quantities. Given a year-frac threshold $\Theta$ that determines the width of the time window, set ${\cal I}_{d} \equiv \lbrace \tilde d: \left | \phi(\tilde d) - \phi(d) \right | \leq \Theta \rbrace$ to be all days of the year that are less than $\Theta$-years away from $d$. The rescaling procedure is as follows---described in continuous variables:
\bitem
	\item For each $\tilde d \in {\cal I}_{d}$ the function
	\begin{align}
H_{\tilde d} (u) := \left ( \frac{\maxG_{d}}{\maxG_{\tilde d}} \right ) \myg_d \left( \langle \hat{s}_{\tilde d}+ u\left(\hat{t}_{\tilde d}- \hat{s}_{\tilde d}\right) \rangle \right), \qquad  u \in [0,1],
\end{align}
	defines a rescaled generation profile for day ${\tilde d}$ with quantities normalized by the ratio of the {maximum} generation of the target day $d$ to that of day ${\tilde d}$.
	\item 
Applying $H_{\tilde d}(\cdot)$ to each of the hour intervals on the target day $d$ yields a set of stationary hourly realizations of rescaled {\it actual} quantities $\tilde{\myg}_{\tilde d}$.
	\item A similar procedure is performed on the forecasts $\myF_d$ to yield rescaled forecasts $\tilde{\myf}_{\tilde d}$.
	\item The maximum hourly generation quantity is estimated as $\maxG_{d,h} := \max_{\tilde d\in{\cal I}_{d}} \left [ \tilde{g}_{\tilde d,h} \right ]$.
	\item {\it For hours $h$ intersecting $\cI_{\tilde d}$} (those with positive probability of generation) production ratio variables are then computed as: $\alpha_{\tilde d,h} = \frac{\tilde{g}_{\tilde d,h}}{\maxG_{d,h}}$; similarly $\beta_{\tilde d,h}$ are computed for the forecasts. Note that the daily vectors $\balph_{d}$ and $\bbet_{d}$ are in $\cR^H$ for $H<24$ and have component values in the unit interval.  Typically $H \sim 10$. The production ratios shown in Figure~\ref{fig:solarBnd} were computed in this fashion.
\eitem

The results presented are based on the rescaling methodology discussed above.
As mentioned, the purpose of the rescaling is to transform actual and realized production data at dates near the target $d$ so as to be consistent with the behavior at the given calibration date--- consistent in the sense that the resulting deviates are (nearly) stationary.  Additional refinements to utilize more information than the max-gen and diurnal boundaries are available but are beyond the scope of this paper and will be described elsewhere.



\subsection{Calibration}

For a given target date $d$ the metacalibration associated with each asset is used to affect the rescaling discussed above yielding a set of production ratios $ \balph_{\tilde d}$ and $\bbet_{\tilde d}$ for actual and forecasted volumes respectively on each date and defined calibration date range $\tilde d \in {\cal I}_{d}$ of width $\Theta$. These are the variables of interest --- the independent variable being the $\bbet$'s, the dependent being the $\balph$'s. 

\begin{figure}[tbph]
\begin{center}
\includegraphics[height=2.5in]{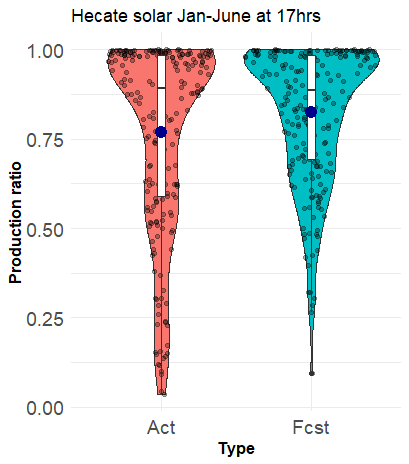}
\includegraphics[height=2.5in]{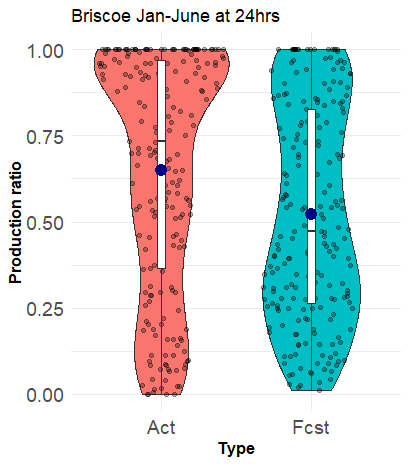}
\caption{\label{Solar-Wind-actFcst} {Realized $\alpha_{d,h}$ vs.~forecasted  $\beta_{d,h}$ production ratios. Left: Solar Asset (Hecate). Right: Wind Asset (Briscoe). The sina violin plots show the 110 days centered on April 1 from 2017 and 2018. The boxplots in the middle provide summary statistics, as well as the respective empirical mean (blue dots), illustrating the bias and different variance of forecasts and actuals.}}
\end{center}
\end{figure}

We observe that the two distributions are qualitatively different and moreover, raw forecast errors are not zero-mean. Figure \ref{Solar-Wind-actFcst} shows the distribution of forecasted $\bbet_{\cdot,h}$ and realized production ratios $\balph_{\cdot,h}$ for a representative wind and solar asset. We fix an hour of the day and consider a window of $110=0.3 \times 365$ days around April 1. Several features are apparent:
\begin{itemize}
  \item Forecasts are biased: $Ave(\myf_{d,h}) \neq Ave(\myg_{d,h})$. This is both intrinsic (for example on a calm day wind forecast might be essentially zero, but realized production is non-negative, so on average will be higher than forecast; an analogous downward bias transpires when forecast production ratios are close to 100\%) and data-driven. It appears that wind forecasts systematically underestimate production.

  \item The distribution of realized production can be bimodal, i.e.~often the production ratios $\balph$ are close to zero or to 100\%. On the other hand, forecast production ratios $\bbet$ are closer to Beta-distributed (and for some assets nearly uniform when tabulated across long periods of time).

  \item The conditional variance of actuals $\alpha_{\cdot,h}$ is much higher for forecasts in the middle than at the edges of the production range (a day forecasted to be cloudless is unlikely to witness materialization of heavy cloud cover). 
\end{itemize}

In order to obtain Gaussianized forecast errors, we proceed to model the conditional mean and variance of $\alpha_{d,h}$ given $\beta_{d,h}$.  The resulting residuals then act as the inputs to the correlation inference within the hierarchical cluster structure. The modeling premise is that for each active hour:
\beqn
\alpha_{\cdot,h} = 0 \vee 1 \wedge  \left[ \mu_h \left( \beta_{\cdot,h} \right ) + \sigma_h \left( \beta_{
\cdot, h} \right ) \cdot Z_{\cdot,h} \right]
\eeqn
with the interpretation being that the realized production ratio $\alpha_{\cdot,h}$ is distributed normally with a mean and variance that depends upon the forecasted production ratio $\beta_{h,\cdot}$, subject to bounds at 0 and 1. Endowing $\sigma_h $ with dependence on $\beta_{\cdot, h}$ is required in order to capture the observed heteroskedasticity; the inclusion of $\mu_h$ and its similar dependence on $\beta_{\cdot, h}$ achieves bias correction and provides a degree of freedom in the calibration method discussed below.

In the results shown in this paper, we have assumed a quadratic form for both  $\mu_h \left( \beta_h \right ) $ and $\sigma_h \left( \beta_h \right )$. To ensure stability of the inference for values of $\beta$ outside of the historical data used in each calibration, the values are first transformed: $\tilde \beta_{d,h} := \psi \left ( \frac {\beta_{d,h}- \mu_{\beta}}{\sigma_\beta} \right )$ where $\psi(x) = \frac{e^x}{1+e^x}$ and the empirical mean and standard deviations of the forecast data are denoted by $\mu_\beta$ and $\sigma_\beta$ respectively. This stabilizes the functional forms of the estimated $\mu_h$ and $\sigma_h$ on the length scale of the empirical standard deviation of the data. Therefore, the functional form used here corresponds to $\mu_h$ and $\sigma_h$ being quadratic in $\tilde \beta_{d,h}$.

Calibration of $\mu_h,\sigma_h$  is via maximum likelihood estimation. The likelihood function is modified appropriately for the values of the argument outside of the unit interval (the point masses). Values of the normal deviates that are not uniquely determined (corresponding to production ratios of 0 or 1) are subsequently inferred by conditional normal calculations using those that are known.
%
Specifically, denoting the forecast error by ${\epsilon}_{d,h} := \alpha_{d,h}-\beta_{d,h}$, and its normalized deviate $ z_{d,h} := \frac{\eps_{d,h} - \mu_h (\beta_{d,h}) }{\sigma_h (\beta_{d,h})}$,
 the coefficients $\vartheta$ defining the parameterizations $\mu_\cdot, \sigma_\cdot$ are set to minimize: 
\begin{align}
{\cal L}\left(\vartheta \right ) & = \sum_{d,h} \left [  {\bf 1}_{\lbrace -\beta_{d,h} < \eps_{d,h} < U_{d,h} \rbrace } \half z_{d,h} (\eps_{d,h})^2+ \log(\sigma_h (\beta_{d,h})) \right ] \nonumber \\
& + \sum_{d,h} \left [  {\bf 1}_{\lbrace \eps_{d,h} \leq  -\beta_{d,h} \rbrace } \frac{1}{F_{\mu_h,\sigma_h} \left ( z_{d,h}(-\beta_{d,h}) \right ) } \int _{-\infty}^{\eps_{d,h}} f_{\mu_h,\sigma_h} (x) \left [  \half z^2_{d,h} (x) + \log (\sigma_h (\beta_{d,h}) ) \right ] dx \right ] \nonumber \\
& + \sum_{d,h} \left [  {\bf 1}_{\lbrace \eps_{d,h} \geq  U_{d,h} \rbrace } \frac{1}{1-F_{\mu_h,\sigma_h} \left ( z_{d,h}(U_{d,h})  \right )  } \int _{U_{d,h}}^{\infty} f_{\mu_h,\sigma_h} (x) \left [ \half z^2_{d,h} (x)  + \log (\sigma_h (\beta_{d,h})  ) \right ] dx \right ]
\end{align}
where $f_{\mu,\sigma}$ and $F_{\mu,\sigma}$ denote  the normal pdf and CDF with mean $\mu$ and variance $\sigma^2$ respectively, and $U_{d,h} = \frac{\maxG_{d}}{\maxG_{d,h}}$ is the level at which asset maximum generation would be exceeded. The first term is the usual form of the normal log-likelihood function and applies to values of $\epsilon_{d,h}$ corresponding to realized production strictly above zero and below maximum generation. The second and third terms are the  expected values of the same likelihood function conditioned on $\epsilon_{d,h}$ breaching the respective bounds.

\begin{figure}[tbph]
\begin{center}
\includegraphics[width=0.32\textwidth]{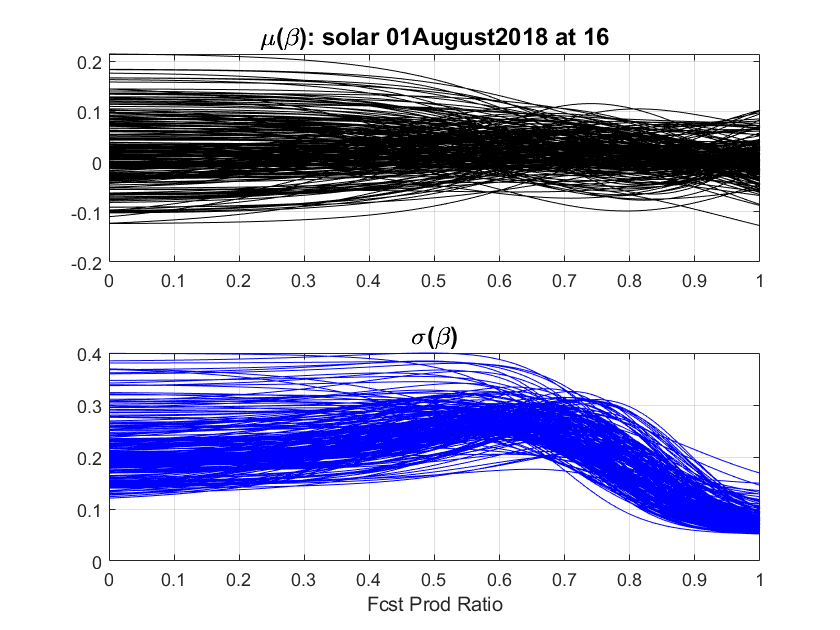}
\includegraphics[width=0.32\textwidth]{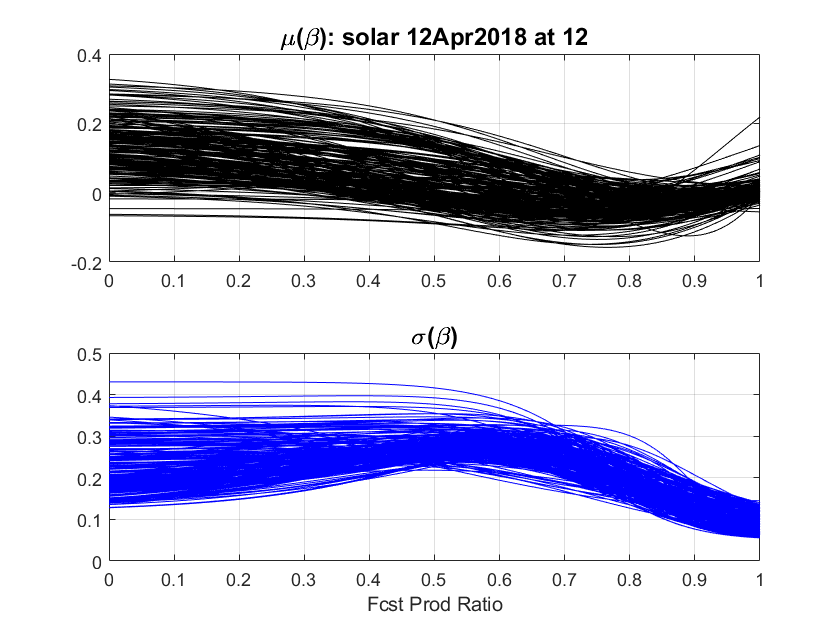}
\includegraphics[width=0.32\textwidth]{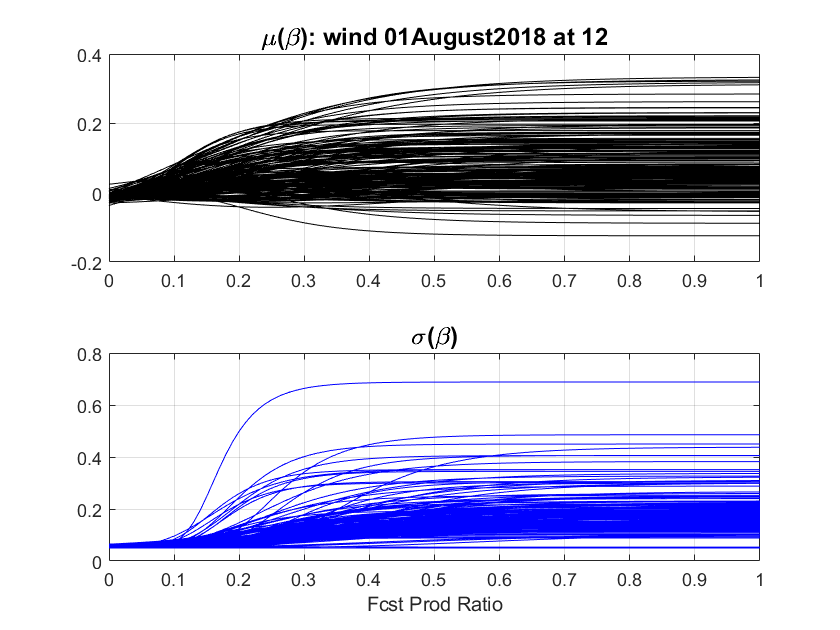} 
\caption{\label{muSigma} {Fitted $\mu_h(\cdot)$ (top row) and $\sigma_h(\cdot)$ across all the assets in the case study for the indicated target date. Left: solar (across 226 assets). Middle: solar (226 assets). Right: wind (264 assets).}}
\end{center}
\end{figure}

An example of such a calibration is shown in Figure \ref{muSigma}. We utilize quadratic fits on the logit of the conditional mean and volatility. Intuitively, $\mu_h(\cdot)$ is the bias correction, translating a forecast production ratio into the expected generation ratio (modulo truncation) and $\sigma_h(\cdot)$ is the heteroskedasticity correction, capturing the higher variability of realized production when forecasts are in the middle of the possible range. As mentioned, the intrinsic constraint that $\alpha_h \in [0,1]$ implies that we expect $\mu_h(0) > 0$ and $\mu_h(1)<0$ and similarly an umbrella shape on $\sigma_h(\cdot)$.

We observe that in our dataset, solar production is typically over-estimated, $\mu_h(\beta ) > 0$ for mid-range forecasts (i.e.~less production tends to materialize on days that are neither very sunny nor very grey), while for wind, production is strongly under-estimated $\mu_h(\beta )< 0$ for mid-range forecasts. This could be partly linked to the concavity/convexity of the respective production curves. For both types of assets, conditional variance $\sigma_h(\beta)$ is highest for forecasts around $\beta=0.6$. For solar, lowest uncertainty $\sigma_h(\beta)$ is on very sunny days ($\beta_{d,h} \simeq 1$), while for wind lowest uncertainty is for very calm days ($\beta_d \simeq 0$) and conditional variances grows in $\beta$. The conditional variance is about the same across solar and wind assets, generally on the scale of $\sigma_{d,h} \in [0.2,0.3]$. All the above are specific to this case study; the overall platform is agnostic to such particularities.

%
%

Each set of realized deviates $z_{d,h}$ is now ``copula-ized" by application of the empirical CDF followed by the inverse of the standard normal CDF. This ensures that cluster construction works in the fully Gaussianized space where Gaussian covariance estimation is correctly specified.
%
%
We will index those hours $h$ for which $\alpha_{d,h} \in (0,1)$, namely not hitting the boundaries of production limits, by ${\mathfrak I}_d$. The calculation of normal deviates used in the subsequent correlation analysis involves the following steps for each active hour.

\begin{enumerate}
	\item Estimate pairwise inter-hour correlations of the normalized residuals $z_{\tilde{d},h} =  \frac{\eps_{\tilde{d},h} - \mu_h (\beta_{\tilde{d},h}) }{\sigma_h (\beta_{d,h})}$ based upon the difference between hours. Specifically assuming that ${\rm corr}\left [ X_{h}, X_{h+k}  \right ] = \rho(k)$ for all hours $h$ and $h+k$ in the set of active hours, calculate the empirical pair-wise correlations using the values computed above. Note that values corresponding to point masses at the boundaries are excluded from these estimates, rendering each estimate for $\rho (k)$ computed from potentially different subsets of the empirical data.
	\item Ensure a positive definite correlation matrix by computing the eigenvalues and eigenvectors for the estimated covariance matrix,  setting all negative eigenvalues to zero, and rebuilding the original matrix, normalizing it by its diagonal elements. 
	\item For each day $\tilde{d}$ in the dataset, assuming joint normality of $\bar X_{\tilde{d}}$, compute the conditional mean and variance for $\bar X_{{\frak I}_{\tilde{d}}^c}$ (these are the deviates that remain unknown since production ratio was zero or 1) given that these were at the boundaries of production limits, given $\bar X_{{\frak I}_{\tilde{d}}}$ which is known. Generate simulations from this joint distribution and for each $h\in{\frak I}_{\tilde{d}}^c$, estimate $x_{\tilde{d},h}$ by the conditional expectation conditioned on exceedance of the boundary of generation production. For example, if the associated $\alpha_{\tilde{d},h}=0$, the estimate is the expected value of the simulations conditioned on $z_{\tilde{d},h}<\frac{-\beta_{\tilde{d},h}-\mu_{h} (\beta_{\tilde{d},h})}{\sigma_{h} (\beta_{\tilde{d},h})}$, this event corresponding to $\alpha_{\tilde{d},h}=0$. A similar condition pertains to the upper boundary.
	\item With a complete set of $z_{\tilde{d},\cot} $ values in hand for the entire dataset, use a standard normal copula by hour to compute normal deviates $\tilde{z}_{\tilde{d},\cdot}$.
\end{enumerate}

\subsection{Simulation }

The last piece of the platform concerns generation of i.i.d.~joint scenarios. Given any target date and a respective asset-level hourly forecast, the platform can output an arbitrary number of joint scenarios. This is done by using the correlation clusters from step (vi) and the historical calibration from step (v). The simulations are generated one by one (vectorized in our code) and start in the  Gaussianized space via conditional normals. At each level of the cluster hierarchy, we have the cluster centroid's deviates being passed down and the cluster-level deviates being then generated from a conditional normal formula. Then we finally reverse the calibration steps to obtain quantities in terms of MWh.

More precisely, we nucleate a multivariate normal sample $E^{(0)}$ for the top-level deviate and then recursively simulate deviates for level-$\ell$ clusters conditional on their ``parent'' $E^{(\ell-1)}$ normal deviates. This is done by conditioning: inputting the already generated $E^{(\ell-1)}$ of the cluster delegate and inferring the rest of the cluster deviates via a conditional Gaussian sampling. Iteration continues to the bottom-most asset level, which yields a complete set of sampled Gaussian $z$'s for each asset and each hour of the day. In the final step, these are inverted through the conditional mean and variance transformations and merged with the forecast $\myf_h$ to yield production simulations in MWh: $([\mu_h(\myf_h) + \sigma_h(\myf_h) E^{(L)}]\cdot M_h + \myf_h) \vee 0 \wedge \maxG_{d,h}$.

%

\section{Results from a Case Study}\label{sec:case-study}

In this section we illustrate our simulation platform with the results from the NREL \texttt{Proposed} testbed. As a start, we generate 1000 scenarios for the fixed day of April 12, 2018. As explained, the scenarios are joint across the 490 assets, and can be understood as 1000 counter-factual realizations of renewable generation on that day, conditional on the given forecast. Since our model captures the cross-asset correlations, one can examine the simulations at any level of generation granularity---marginal at each asset, aggregated by a geographic region, aggregated at a zonal level, or across the entire grid. The latter options just require summing up across asset subsets by scenario. Similarly, since the model captures the temporal correlations across hours, we can sum up across periods to obtain scenarios for aggregated daily generation, etc.

\begin{figure}[tbph]
\begin{center}
\includegraphics[height=2.5in]{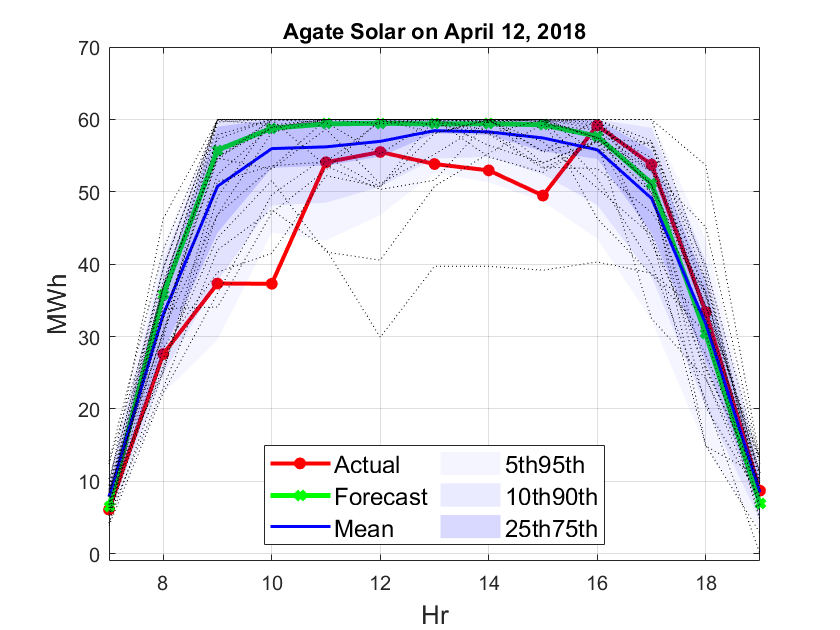}
\includegraphics[height=2.5in]{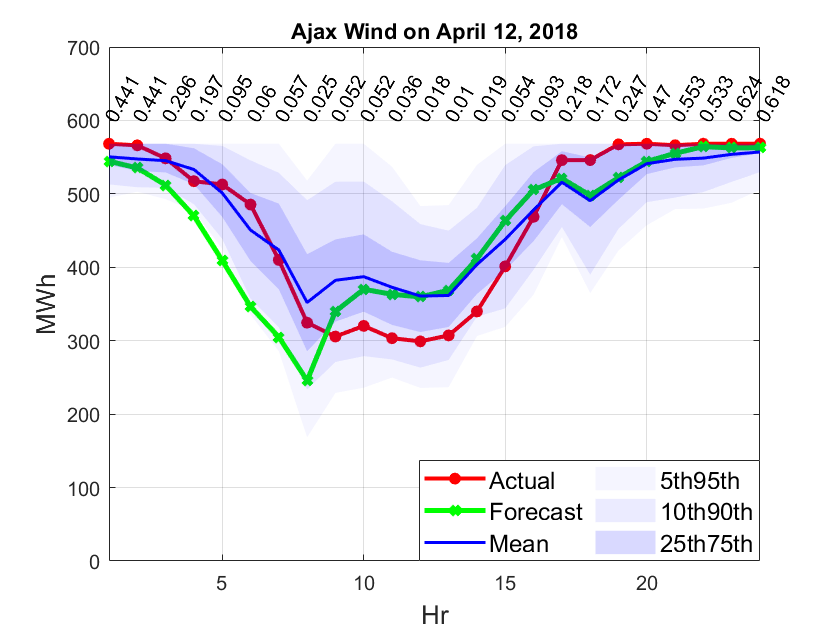} 
\caption{\label{Solar-Wind-Sim} {Representative asset-level scenarios. Left: Solar Asset (Agate) with 20 daily scenarios. Right: Wind Asset (Ajax). Numbers indicate the probability of the point masses for maximum wind generation. All simulations are calibrated to 04/12/2018. }}
\end{center}
\end{figure}

  Figure \ref{Solar-Wind-Sim} shows the hourly-based view of single asset simulations, for a solar (Castro) and a wind asset (Aguayo) respectively. The Figure shows the forecast $f_h$, the realization $g_h$ and the mean simulated generation $m_h=Ave(g^{1:1000}_h)$ based on 1000 scenarios $g^{1:1000}_h$. We also show the 95\%-scenario band, obtained by sorting for each hour $g^i_h$ and then saving the respective 2.5\% and 97.5\% quantiles $q^{\alpha}_h := ( g^{1:1000}_h)$. We observe that $m_h \neq \myf_h$ per the de-biasing calibration based on $\mu_h(\cdot)$; we also observe the non-constant standard deviation $\hat{\sigma}_h = StDev (g^{1:1000}_h)$ that reflects the non-constant conditional variance $\sigma_h(\beta)$ and the temporal pattern of $\myf_{1:24}$.

As mentioned, there are often non-zero probabilities of zero or maximum generation, for example on the right panel of Figure \ref{Solar-Wind-Sim} maximum wind production has about 50-69\% chance of occurrence in the late evening, 

Figure \ref{fig:zonalSims} in the Appendix repeats the above for the aggregated generation in the Far West zone that contains 46 solar assets and 61 wind assets. We observe some partial diversification with tighter relative uncertainty bands, but still quite a bit of variability at the zonal level.
%
%
Figure \ref{fig:hrCorr} in the Appendix shows the realized intra-day hourly correlations for representative solar and wind assets. We observe decorrelation taking hold after 4-6 hours.

\subsection{Probabilistic Assessment}\label{sec:assess}

To assess the generated simulations, the main tool is to compare the (randomly sampled) scenario distribution $F_t(\cdot)$ to the realized actual $g_t$. This implies defining a loss metric $D(F,g)$ and then averaging it across a set of test hours or day  \cite{lauret2019verification,woodruff2018constructing,rachunok2020assessment,ziel2018probabilistic}. The metric judges the closeness of the realized actual to the distribution; the averaging is necessary to draw statistics about the distribution of the forecast fitness. We note that the oft-mentioned nonstationarity means that each $F$ and $g$ come from different underlying distributions, hence the averaging is also important to average out performance across different potential settings (i.e.~months of the year, different weather patterns, etc). To that extent, one should not draw any conclusions from performance on a given test instance, and focus on aggregate performance.

\begin{figure}[tbph]
\begin{center}
\includegraphics[height=2in,trim=1in 3.2in 1in 3.2in,clip=true]{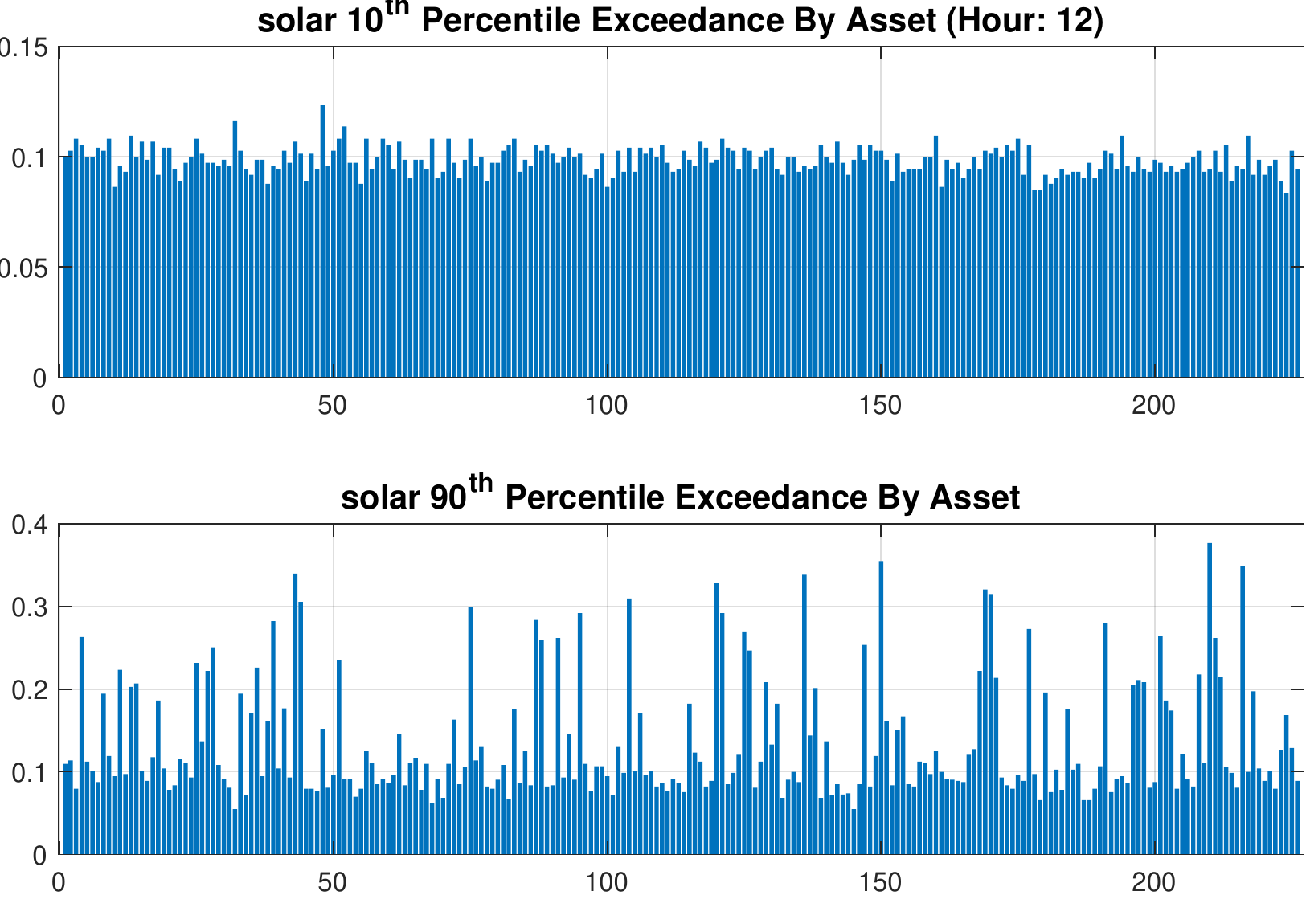} 
\includegraphics[height=2in,trim=1in 3.2in 1in 3.2in,clip=true]{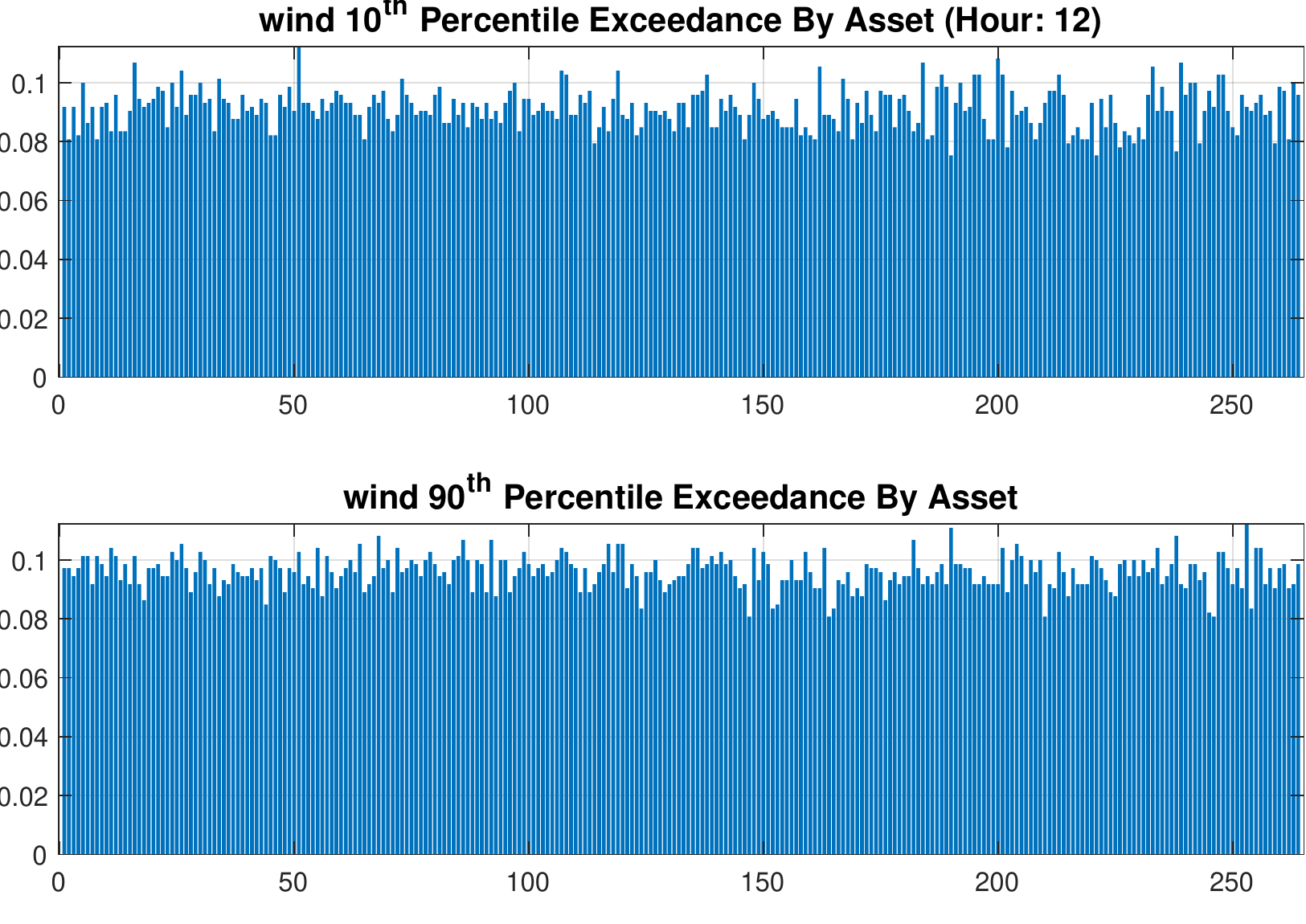} 
\caption{{\label{fig:WindStats} Summary of scenario coverage probabilities by asset. Left: 226 solar assets. Right: 264 wind assets. Both panels are for noon $h=12$ across the entire year 2018. We show $\sum_{d=1}^{365} 1_\{g_d < q_{0.1}\}$ (top) and $\sum_{d=1}^{365} 1_\{g_d > q_{0.9}\}$ (bottom), i.e.~the frequency of the actuals being in the left/right tail of the scenario distribution.  }}
\end{center}
\end{figure}

The Probability Integral Transform (PIT) looks at the realized percentile, $F_t(g_t)$. Under the hypothesis that $F$ perfectly captures the actuals, $F_t(g_t) \sim U(0,1)$ should be uniform, hence one may test the uniformity of the PIT. More locally, one may test statistical coverage, i.e.~the frequency that a particular range of percentiles appear; this is especially relevant for checking the tails, for example how frequently are the actuals far from the bulk of the simulations (i.e.~yield extreme percentiles close to zero or to 1). 
Figure~\ref{fig:WindStats} shows summaries of coverage for the 10\% and 90\% quantiles across all the assets. Since the nominal coverage level is chosen to be 10\% in both cases, under a perfect statistical fit, the exceedance probabilities would be close to 10\% in all 4 panels. We observe that this is indeed so in 3 of the 4 panels; for some solar assets we observe a higher frequency of very high generation (i.e.~scenarios underestimating the probability of sunny conditions).

More generally, one may consider the so-called strict scoring rules  \cite{gneiting2007strictly,gneiting2007probabilistic,gneiting2014probabilistic}, such as the Continuous Ranked Probability Score (CRPS) for univariate assessment and the Energy Score (ES) for multivariate assessment. 




\begin{figure}[tbph]
\begin{center}
\includegraphics[height=2in,width=5in]{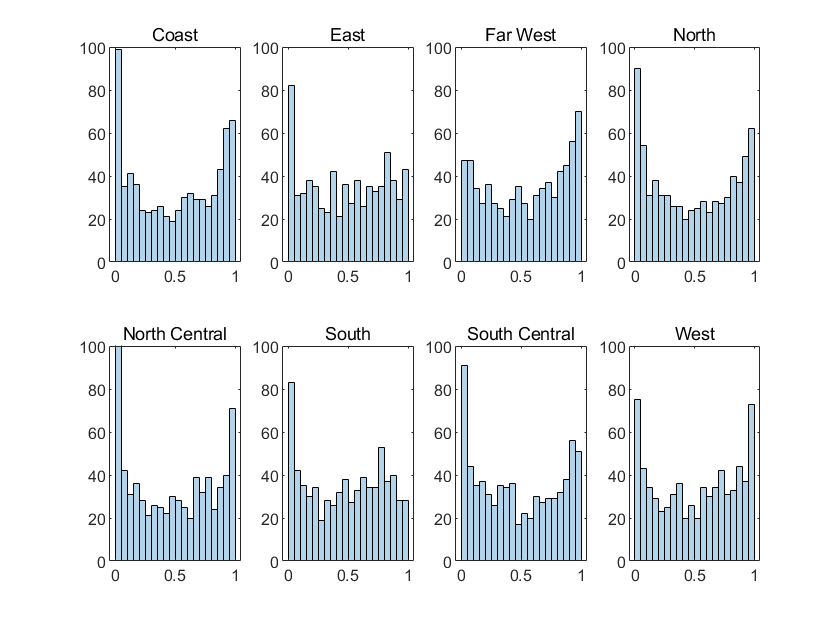}
\caption{{\label{fig:windZonalPit} PIT histograms for zonal wind generation. We show the percentiles for the 30 days in April 2018 and all 24 hours of the day (720 total hours).}}
\end{center}
\end{figure}

Figure~\ref{fig:windZonalPit} shows the PIT histograms for the aggregated zonal wind production for each of the 8 ERCOT zones. We observe that the histograms are very close to uniform except for some exceedances at the lowest decile (in other words, there is more than expected frequency of days/hours where actuals are substantially below all scenarios). This is a material consistency check since the distribution of the sum is sensitive to the correlation structure, and moreover zonal production is important for SCUC and SCED.

\section{Conclusion}\label{sec:conclude}

The presented platform provides a novel framework for generating day-ahead scenarios for short-term grid operational planning. The outputted simulations can be used for uncertainty quantification at multiple stages of the daily ISO tasks: for stochastic optimization during security-constrained unit commitment, for risk indexing to rank renewable assets (which  otherwise all have zero marginal generation costs), for risk planning to anticipate reserves needs, etc. Moreover, our probabilistic framework is amenable to additional extensions. For example, we have implemented an extension for intra-day simulations: generating hourly scenarios for 6-hour blocks conditioned on $T-6$ forecast updates. Since the methodology is agnostic to many of the empirical features, it can also be modified to provide scenarios at the sub-hourly scale (e.g.~15- or 5-min intervals). The method can also be applied modulo minor adjustments to load modeling; the latter is nowadays highly stochastic due to large amounts of behind-the-meter rooftop photovoltaic panels which induce correlation between load and solar.

Several aspects of our model warrant further investigation, especially in terms of the correlation modeling. The approach in Section~\ref{sec:clustering} works solely with the empirical normalized deviates, and is not aware of any spatial structure. Given that generation is weather-driven, incorporation of spatial constraints, for instance to ensure high correlation of closely located assets could be considered. Similarly, our approach concentrates on creating a hierarchical structure of the covariance matrix; other regularization, such as sparse precision matrices are alternatives to be analyzed. On the calibration side, all described calibration steps are currently done asset-by-asset. Information fusion, for example to improve estimation of hourly maximum or mean generation via a Bayesian framework, could be beneficial. We continue to actively develop the platform and several of the above extensions will be addressed in subsequent articles.

\subsubsection*{Acknowledgements:}  Both authors are partially supported via the ARPA-E PERFORM grant DE-AR0001289. We are thankful to Rene Carmona, Xinshuo Yang, Arvind Shrivats and Mahashweta Patra for many useful discussions. AS and MP have also contributed some of the code for probabilistic assessment. We also thank ARPA-E and the PERFORM Data Plan teams (especially Texas A\&M, NREL and Wisconsin) for providing the dataset used for illustration throughout the article. 

\bibliographystyle{alpha}
\normalfont\small 
\bibliography{orfeus}

\section*{Appendix}

\begin{figure}[tbph]
\begin{center}
\includegraphics[height=2.5in]{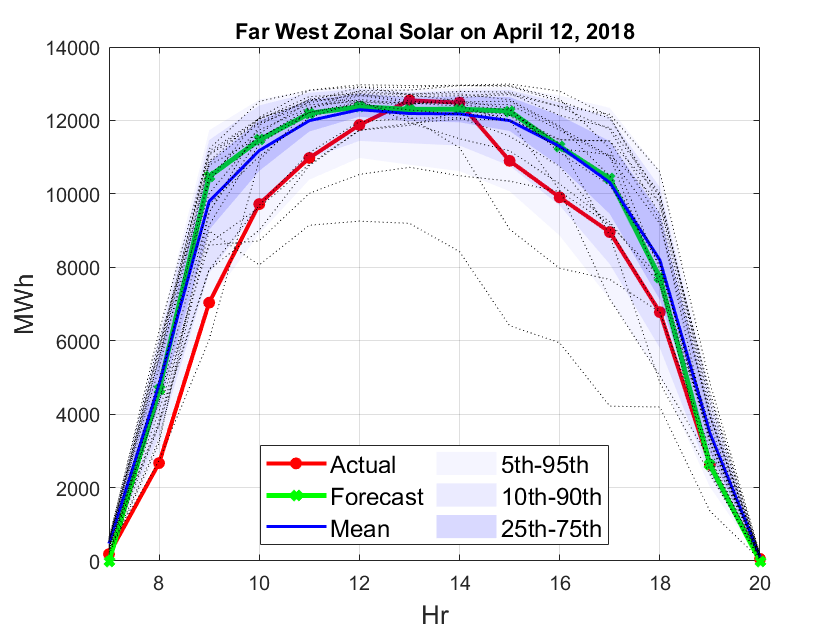}
\includegraphics[height=2.5in]{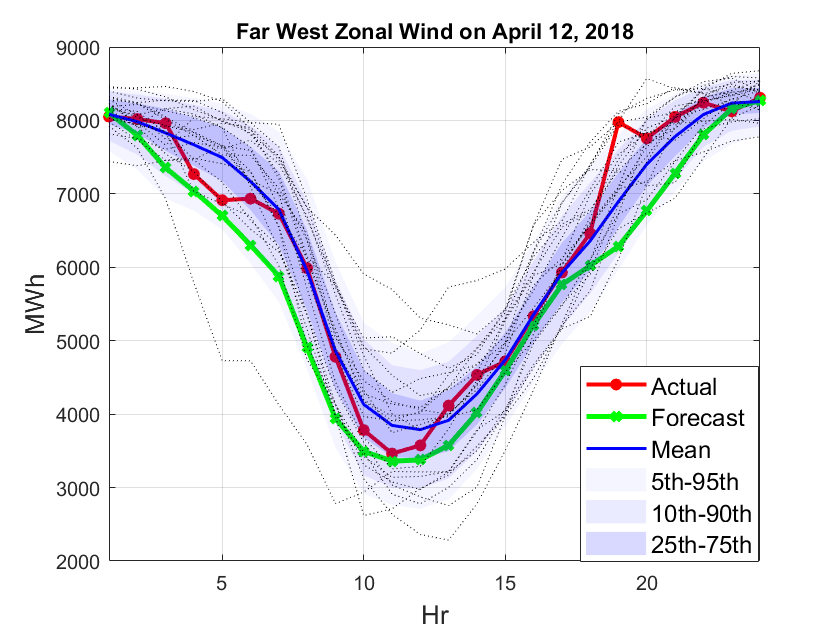} 
\caption{\label{fig:zonalSims} {Representative zone-level scenarios. Left: Solar in Far West (61 assets); Right: Wind in Far West (54 assets). All simulations are calibrated to 04/12/2018. }}
\end{center}
\end{figure}

\begin{figure}[tbph]
\begin{center}
\includegraphics[width=0.4\textwidth]{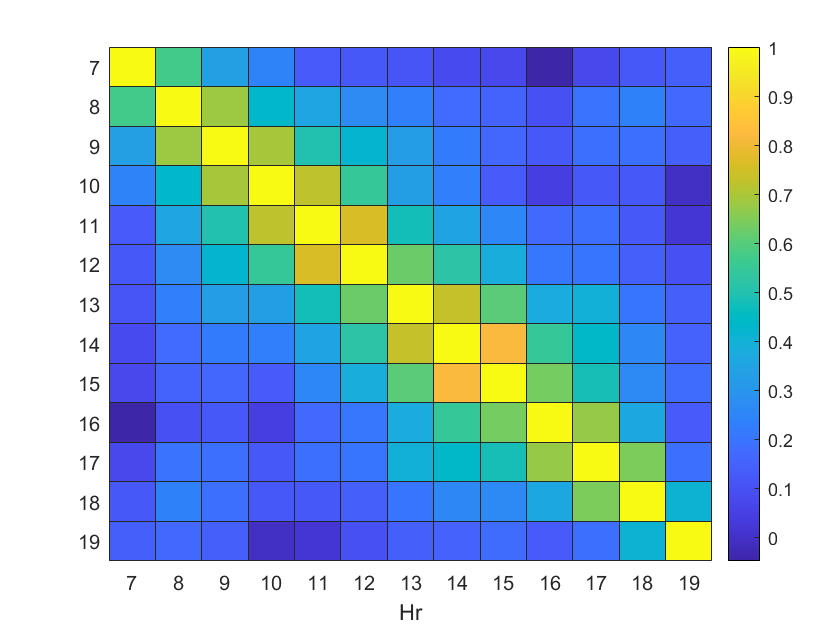}
\includegraphics[width=0.4\textwidth]{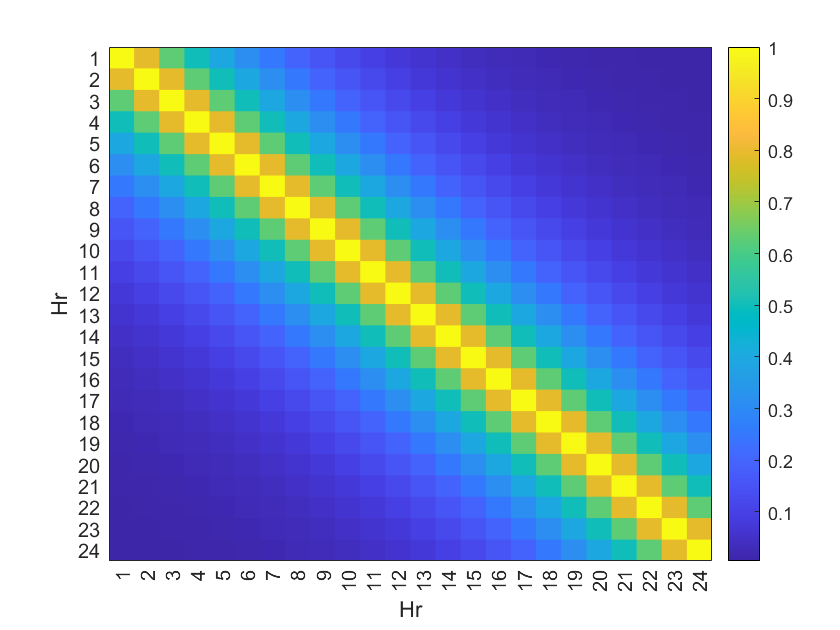}
\caption{\label{fig:hrCorr} {Empirical correlation matrix across the active hours of April 13, 2018. Left: Adamstown solar (13 active hours). Right: Amazon wind farm.}}
\end{center}
\end{figure}

\end{document}